\documentclass[a4paper,11pt]{article}	% Choose this for a physics paper
\pdfoutput=1 % if your are submitting a pdflatex to arXiv
\usepackage{jheppub} % If you decide to use the JHEP style
\usepackage{colortbl}
\usepackage{booktabs}                          % better tables: \toprule,\midrule,bottomerule
\usepackage{multirow,bigdelim}                 % things span over rows of a table: \rdelm, \multirow

\usepackage{units}                             % \unit[val]{dim}, \unitfrac[val]{num}{denom}, \nicefrac[fontcmd]{num}{denom}
\usepackage{amsmath,amssymb,mathtools}         % to enable some mathematical symbols	
\usepackage{slashed}                           % for p slash
\usepackage{mathrsfs}                          % for scripture font, \mathscr
\usepackage{amsthm}
\usepackage{bbm,dsfont}                        % provide \mathbbm{1}(serif) and \mathds{1}(san serif) respectively

\usepackage{bm}                                % provide \bm{} for bold greek letters and other math symbols
\usepackage{upgreek}                           % upright normal greek letters \upalpha\Upalpha

\usepackage{graphicx}                          % to enable the inserting of graphics
\usepackage{subfig}	                           % for \subfloat commands
\usepackage{tikz}                              % the almighty PGF/TikZ! Useful TikZ packages: positioning,scopes,chains,matrix,arrows,calc,trees,intersections.
\usepackage{float}
\usepackage[export]{adjustbox}                 % align figures
\usepackage{url}                               % for url reference
\usepackage{makeidx} % macro package for index making
\usepackage[numbers,sort&compress]{natbib} % bibliography
\usepackage{ifthen}	                           % enables \ifthenelse{condition}{command1}{command2}
\usepackage{makeidx}

\usetikzlibrary{calc,positioning, intersections, arrows, shapes, fit,
  tikzmark, decorations.markings}

\def\mc{\mathcal}
\def\md{\mathbf}
\def\mf{\mathfrak}
\def\ms{\mathsf}
\def\mr{\mathscr}

\newcommand{\wt}[1]{\widetilde{#1}}

\newcommand{\pd}{\partial}

\def\vev#1{\left\langle #1 \right\rangle}

\DeclarePairedDelimiterX\opm[3]{\langle}{\rangle}{#1 \delimsize\vert #2 \delimsize\vert #3}
\DeclarePairedDelimiterX\ip[2]{\langle}{\rangle}{#1 \delimsize\vert #2}
\def\IC{\mathbb{C}}

\def\IF{\mathbb{F}}

\def\IP{\mathbb{P}}

\def\IR{\mathbb{R}}
\def\IZ{\mathbb{Z}}

\DeclareMathOperator{\Li}{Li}

\DeclareMathOperator{\real}{Re}
\DeclareMathOperator{\imag}{Im}

\newcommand{\nn}{\nonumber \\}

\newcommand{\re}{{\rm e}}
\newcommand{\ri}{{\mathsf{i}}}
\newcommand{\rd}{{\rm d}}

\newcommand{\CA}{\mc{A}}
\newcommand{\CD}{\mc{D}}
\newcommand{\CF}{\mc{F}}
\newcommand{\CO}{\mc{O}}
\newcommand{\CP}{\mc{P}}
\newcommand{\CN}{\mc{N}}
\newcommand{\CS}{\mc{S}}
\newcommand{\CT}{\mc{T}}
\newcommand{\CX}{\mc{X}}
\newcommand{\CZ}{\mc{Z}} 

\newcommand{\SD}{\ms{D}}

\newcommand{\disc}{\text{disc}}

\theoremstyle{plain}                       % default

\theoremstyle{definition}

\newtheorem{coj}{Conjecture}[section]

\newenvironment{GMNrevision}
  {\par\begingroup\color{red}%
   \hypersetup{citecolor=red,linkcolor=red,urlcolor=red}}
  {\par\endgroup}

\graphicspath{{figs/}}

\makeindex

\title{\boldmath BPS invariants from resurgent topological string}
\author[a]{Jie Gu}
\affiliation[a]{School of Physics and Shing-Tung Yau Center\\
  Southeast University, Nanjing 210096, China}%

\emailAdd{jie-gu@seu.edu.cn}

\abstract{We review recent progress on the conjectural relation
  between the resurgent structure of topological string free energies
  and BPS invariants of type II string theory compactified on the same
  Calabi--Yau threefold. In this picture, Borel singularities are
  located at integral periods associated with central charges of
  stable D-brane bound states, while Stokes constants encode their BPS
  indices, or generalized DT invariants. After introducing resurgence
  and perturbative topological strings, we explain how trans-series
  solutions of the holomorphic anomaly equations determine
  non-perturbative amplitudes and illustrate the proposed
  correspondence in local and compact geometries. We discuss the
  behavior under changes of polarization and moduli, and review
  supporting arguments based on exact WKB analysis, blowup equations,
  the Stokes automorphisms of flat coordinates and dual partition
  functions, and Kontsevich--Soibelman wall-crossing. We conclude with
  open questions concerning a general proof, the interpretation of
  non-perturbative amplitudes, and the construction of
  non-perturbative completions.}


\begin{document}
\maketitle
\flushbottom

\section{Introduction}
\label{sc:intro}

String theory is usually defined only perturbatively.  For instance,
the free energy in string theory is given by a perturbative series in 
the string coupling $g_s$,
\begin{equation}
  \CF(g_s) = \sum_{g=0}^\infty \CF_g g_s^{2g-2},
\end{equation}
where each coefficient $\CF_g$ is defined individually by an
integral over the moduli space of genus-$g$ Riemann surfaces
describing string worldsheets.  Since the string coupling $g_s$ is
determined by the exponential of the background value of the dilaton
\cite{Polchinski:1998rq,Polchinski:1998rr}, it is meaningful to sum
the perturbative series and consider the total free energy in the
presence of the dilaton field.  However, the string free energy is
divergent for any non-vanishing value of $g_s$, as the coefficients
$\CF_g$ grow factorially with the genus $g$
\cite{Gross:1988ib,Shenker:1990uf}
\begin{equation}
  \CF_g \sim (2g)!.
\end{equation}
The divergence of the perturbative series indicates that the
perturbative series by itself is not the complete description, and
there must be non-perturbative contributions of the order
$\CO(\re^{-1/g_s})$.  The natural questions are then: what are the
physical origins of these non-perturbative contributions, and how can
they be calculated?  Similar questions arise in quantum mechanics, for
instance, in studying energies of stationary states in the
semiclassical limit.  The perturbative series of the energy eigenvalues
are divergent, and they must be supplemented by non-perturbative
contributions.  The latter are known to arise due to physical
phenomena such as quantum tunneling induced by instanton effects, and
they can be calculated by either a saddle-point expansion of
the Euclidean path integral around an instanton or the exact
Wentzel--Kramers--Brillouin (WKB) method.

In string theory, Dirichlet branes (D-branes) provide a natural
candidate for these effects. Polchinski showed that Dirichlet
boundaries produce effects with the characteristic exponential scale
$\CO(\re^{-1/g_s})$ \cite{Polchinski:1994boundary}, and D-branes,
which provide these boundary conditions, are natural ingredients in
string theory \cite{Polchinski:1995mt}.  Checking the detailed
relation between D-brane states and the non-perturbative sectors of a
given string amplitude remains difficult.

It is therefore natural to consider a simple version of string theory
to test this idea.  A particularly useful candidate is
topological string theory.  Consider topological string theory
whose target space is a Calabi--Yau threefold $X$ with varying K\"ahler
moduli $\md{t}$.  The perturbative free energy of topological string
theory is
\begin{equation}
  \CF(\md{t},g_s) = \sum_{g=0}^\infty \CF_g(\md{t})g_s^{2g-2},
\end{equation}
where the coefficients $\CF_g(\md{t})$ depend on the K\"ahler moduli
$\md{t}$.  In contrast to superstring theory, the topological string
free energy has a rigorous mathematical definition.  The free energies
$\CF_g(\md{t})$ are the generating series of the Gromov--Witten
invariants $N_{g,\md{d}}$ \cite{Bershadsky:1993cx}, which count
stable pseudoholomorphic maps from genus-$g$ Riemann surfaces in a
curve class $\md{d}$ in the Calabi--Yau threefold $X$,
\begin{equation}
  \CF_g(\md{t}) = \sum_{\md{d}\in H_2(X)} N_{g,\md{d}} \re^{-\md{t}\cdot\md{d}}.
\end{equation}
From a physical point of view, the free energies $\CF_g$ with
$g\geq 1$ calculate the coefficients of the F-term contribution in
the effective action of type IIA supergravity compactified on $X$
\cite{Bershadsky:1993cx}
\begin{equation}
  S_{\text{eff}} \supset \int \rd^4 x \CF_g(\md{t}) R_+^2 F_+^{2g-2},
\end{equation}
where $R_+$ is the self-dual Riemann tensor and $F_+$ the
self-dual field strength of the graviphoton field in the same Weyl
supermultiplet.  Furthermore, in topological string theory, powerful
computational methods such as the holomorphic anomaly equations
\cite{Bershadsky:1993cx} allow the free energies to be calculated
efficiently.  In some quintessential topological string models on
non-compact Calabi--Yau threefolds, one can calculate hundreds of
terms of the free energy, and even in compact models such as the
quintic, it is now also possible to calculate more than sixty terms of
the perturbative free energy \cite{Alexandrov:2023zjb}.  This provides
extensive perturbative data for non-perturbative analysis.  But before
turning to a non-perturbative analysis, we need to clarify what kind
of questions we wish to address.

In discussing non-perturbative contributions to an observable, there
are typically two kinds of questions.  First, one can calculate the
exact observable, including all non-perturbative
corrections, provided that a non-perturbative definition of the
observable as a function is available.  Second, if such a definition
is not available, one can also ask which
non-perturbative contributions can supplement the perturbative series
on physical or mathematical grounds.  In string theory, a
non-perturbative definition is very rare\footnote{For
  topological string theory, a non-perturbative definition is provided
  by the topological string/spectral theory (TS/ST) correspondence
  \cite{Grassi:2014zfa,Codesido:2015dia,Marino:2015nla,Marino:2024tbx}. But
  this is not the focus of this paper.}, so we focus on
the second question in topological string theory.

\begin{figure}
  \centering%
  \includegraphics[height=3.5cm]{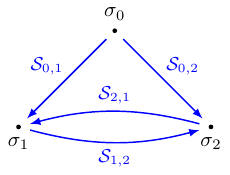}%
  \caption{Stokes automorphism}%
  \label{fig:Stokes-automorphisms}%
\end{figure}

For this purpose, we can exploit a powerful mathematical theory called
resurgence \cite{Ecalle,Sauzin:2014intro}. Resurgence theory is the
mathematical framework developed by the French mathematician Jean
Ecalle to study the relationship between asymptotic series and
non-perturbative contributions.  Resurgence theory has three important
messages that are particularly pertinent for us.  First, given
a divergent power series
\begin{equation}
  \varphi^{(0)}(z) = \sum_{n=0}^\infty a_n z^n,
\end{equation}
with asymptotic behavior of the coefficients
\begin{equation}
  a_n \sim n!,
\end{equation}
possible non-perturbative contributions come in the form of
trans-series
\begin{equation}
  \varphi(z) = \varphi^{(0)}(z) + \sum_{i\geq 1} \sigma_i \varphi^{(i)}(z),
\end{equation}
where $\varphi^{(i)}(z)$ is of the form
\begin{equation}
  \varphi^{(i)}(z) = \re^{-\CA_i/z} \sum_{n} a_n^{(i)} z^n,
\end{equation}
with a non-perturbative exponential suppression factor and an
additional power series.  Second, the trans-series can be promoted to
an analytic function by means of Borel resummation
\begin{equation}
  \varphi(z) \rightarrow \phi(z) = \mr{S}\varphi(z).
\end{equation}
Finally, $\phi(z)$ is naturally defined as an analytic function of
the complex variable $z$. To obtain consistent analytic
continuations over the entire complex plane, the non-perturbative
contributions $\varphi^{(i)}(z)$ and the perturbative series
$\varphi^{(0)}(z)$ must be related by linear transformations called
Stokes automorphisms, as illustrated by
Fig.~\ref{fig:Stokes-automorphisms}.  With the help of Stokes
automorphisms, it is possible to extract non-perturbative
contributions from the perturbative series, or in other words, the
non-perturbative contributions resurge from the perturbative series,
hence the name resurgence.

Given that long perturbative series can be calculated for topological
string free energy, it is natural to apply resurgence theory to study
the non-perturbative contributions.  This program began at the start
of the twenty-first century
\cite{Marino:2006hs,Marino:2007te,Marino:2008ya}.  Over the years,
many results have accumulated
\cite{Pasquetti2010,Aniceto2012,Couso-Santamaria2016,Couso-Santamaria2015,Grassi:2019coc,Gu:2021ize,Alim:2021mhp,Grassi:2022zuk,Gu:2022sqc,Gu:2023mgf,Gu:2023wum,Iwaki:2023rst,Alexandrov:2023wdj,Marino:2024yme,Gu:2024wag,Douaud:2024khu,Douaud:2026qfo,Guo:2026xpg,Marino:2024tbx},
and they have led to the conjecture that non-perturbative
contributions to topological string free energy correspond to stable
D-brane bound states in type II superstring theory.  More concretely,
each non-perturbative contribution in the form of trans-series
corresponds to a stable D-brane bound state, or BPS state. The
non-perturbative action $\CA$ is the central charge of the bound
state, while the Stokes constant relating this non-perturbative
contribution to the perturbative free energy gives the multiplicity of
the bound state, known as a BPS invariant in the physics literature
and a generalized Donaldson--Thomas (DT) invariant in the mathematics
literature.
Substantial non-trivial numerical evidence and various physical
arguments with differing levels of rigor have emerged, leading to
important applications.  It is the purpose of this article to review
this exciting progress.

% \begin{Historyrevision}
% The availability of long perturbative series makes topological string
% free energies a natural setting for resurgence. Beginning with open
% strings and matrix models \cite{Marino:2006hs,Marino:2007te,Marino:2008ya},
% this approach led to trans-series solutions of the holomorphic anomaly
% equations \cite{Couso-Santamaria2016,Couso-Santamaria2015} and to a
% proposed identification of primitive Stokes constants with
% Bogomolny--Prasad--Sommerfield (BPS) invariants, or generalized
% Donaldson--Thomas (DT) invariants
% \cite{Gu:2021ize,Iwaki:2023rst,Marino:2024tbx}.
% Section~\ref{sc:np-TS} traces the development of the three ingredients
% of this proposal: integral-period actions, non-perturbative amplitudes,
% and BPS Stokes data.
% \end{Historyrevision}

% More concretely, each non-perturbative contribution in the form of
% trans-series corresponds to a stable D-brane bound state, or BPS
% state. The non-perturbative action $\CA$ is the central charge of
% the bound state, while the Stokes constant relating this
% non-perturbative contribution to the perturbative free energy gives
% the multiplicity of the bound state, known as a BPS invariant in the
% physics literature and a generalized DT invariant in the mathematics
% literature.
% %
% Substantial non-trivial numerical evidence and various physical
% arguments with differing levels of rigor have emerged, leading to
% important applications.  It is the purpose of this article to review
% this exciting progress.

\section*{Acknowledgement}

This is an invited submission to a special issue of Advances in
Theoretical and Mathematical Physics as proccedings to the
``Calabi-Yau 50 conference'' in 2026.  The author is grateful for the
opportunity to review his work on this special occasion and sincerely
thanks his many wonderful collaborators.  The author is supported by
National Natural Science Foundation of China through Grant
No. 12375062.

\section{Resurgence}
\label{sc:rs}

In physics, perturbative series such as
\begin{equation}
  \varphi(z) = \sum_{n=0}^\infty a_n z^{n+1},
  \label{eq:varphi}
\end{equation}
are often asymptotic series of Gevrey order one, which means that their
coefficients grow factorially,
\begin{equation}
  a_n \sim n!,
\end{equation}
so that the series has zero radius of convergence.  Thus, however
small a nonzero value of $z$ is, the power series does not converge.

To deal with such asymptotic series, it is beneficial to
recall that many of these series arise from a Gaussian expansion of a
finite- or infinite-dimensional exponential integral with a
holomorphic action $S(t)$ near a saddle point $t_\star$
\begin{equation}
  \phi(z) = \int_\Gamma \rd t \re^{-S(t)/z} \sim \varphi(z).
\end{equation}
On the other hand, in the complex plane of the integration variable
$t$, one can attach to the saddle point $t_\star$ a Lefschetz thimble
$\Gamma_\star$ or stable gradient flow with the property
\begin{equation}
  \imag \frac{S(t) - S(t_\star)}{z} = 0,\quad \real \frac{S(t)-S(t_\star)}{z} >0,
\end{equation}
such that integration along $\Gamma_\star$ is convergent.
This suggests that the asymptotic series associated
with the saddle point can be promoted to a complex function, defined by
integration along the Lefschetz thimble attached to the same
saddle point.  This can sometimes be done by the method of Borel
resummation.

Given a Gevrey-order-one series \eqref{eq:varphi}, as its divergence is due to
the factorial growth of its coefficients, we can cure its divergence
by dividing its coefficients by factorials.  In this way, we construct
an auxiliary series known as the \emph{Borel transform} from
\eqref{eq:varphi},
\begin{equation}
  \mr{B}\varphi(\zeta) = \sum_{n=0}^\infty \frac{a_n}{n!}\zeta^n,
\end{equation}
which has a positive radius of convergence.  The radius of convergence
is bounded by singular points.  Suppose these singular points are not
dense and that the Borel transform can be analytically continued to
infinity along a direction $\theta$ with at most
exponential growth, i.e. there exist constants $K,R > 0$ such that 
\begin{equation}
  \left|\mr{B}\varphi(\zeta)\right| < K \re^{|\zeta|/R},\quad \forall\zeta \in \re^{\ri\theta}\IR_+,
\end{equation}
then the Laplace transform of the Borel transform, 
\begin{equation}
  \mr{S}^{(\theta)}\varphi(z) =
  \int_0^{\re^{\ri\theta}\infty}\rd\zeta \re^{-\zeta/z}\mr{B}\varphi(\zeta),
\end{equation}
known as the \emph{Borel resummation}, is convergent and well-defined
for $z$ in the domain
\begin{align}
  z \in D^{(\theta)}_R
  &= \{ z\in\IC\,|\, \real(\re^{\ri\theta}/z) > 1/R\} \nn
  &= \{ z=x+\ri y\in\IC\,|\, (x-R/2\cos\theta)^2+(y-R/2\sin\theta)^2 < (R/2)^2\}.
\end{align}
If the Borel transform has at most polynomial growth, we may take
$R\rightarrow\infty$, and the Borel resummation converges for
\begin{align}
  z \in \lim_{R\rightarrow\infty} D^{(\theta)}_R = D^{(\theta)} = \{ z\in\IC \,|\, |\arg (z) -\theta | < \pi/2\}.
\end{align}
It is easy to check that the Borel resummation admits $\varphi(z)$ as
its asymptotic expansion for $z \in D^{(\theta)}_R$.  Furthermore, if
$\varphi(z)$ does arise from a Gaussian expansion of a finite-
dimensional exponential integral $\phi(z)$ near a non-degenerate
saddle $t_\star$, it can be shown that the Borel resummation equals
integration along the Lefschetz thimble $\Gamma_\star$ attached to
the saddle point $t_\star$
\cite{Malgrange1980,Pham:1983van,Pham1985,berry:1991hyper,howls:1997hyper,delabaere:2002global}.

\begin{figure}
  \centering \subfloat[lateral
  resummation]{\includegraphics[height=4.5cm]{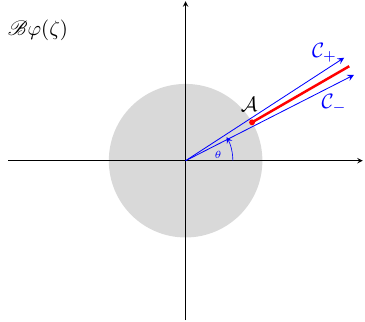}\label{fig:lateral}}%
  \hspace{5ex} \subfloat[Stokes
  discontinuity]{\includegraphics[height=4.5cm]{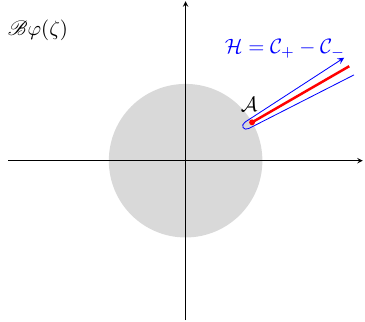}\label{fig:disc}}%
  \caption{Borel singularity and Stokes discontinuity}
  \label{fig:brlplane}
\end{figure}

The definition of Borel resummation hinges upon the possibility of
analytically continuing the Borel transform to infinity along a given
direction.
Therefore, the analytic properties of the Borel transform in the
complex $\zeta$ plane, also known as the \emph{Borel plane}, are
very important.
Analytic continuation of the Borel transform of a Gevrey-order-one
series may encounter singularities.  A ray in the Borel plane from the
origin to infinity that passes through a singular point is called a
\emph{Stokes ray}.  Along the direction of a Stokes ray, the
integration in the definition of Borel resummation will be obstructed.
One can bypass the singular point by tilting the integration contour
slightly above or below the singular point.  In this case, we do not
have a unique Borel resummation, but a pair of lateral resummations,
\begin{equation}
  \mr{S}^{(\theta_\pm)}\varphi(z) =
  \int_0^{\re^{\ri(\theta\pm\epsilon)}\infty}
  \rd\zeta \re^{-\zeta/z}\mr{B}\varphi(\zeta),
\end{equation}
as shown in Fig.~\ref{fig:lateral}.  Their difference, known as the
\emph{Stokes discontinuity}, can be calculated by integration along a
contour of Hankel type that goes around the singular point, as in
Fig.~\ref{fig:disc}, and its value is asymptotically
\begin{equation}
  \disc_\theta\varphi(z) := \mr{S}^{(\theta_+)}\varphi(z) -
  \mr{S}^{(\theta_-)}\varphi(z) \sim
  z^{-\alpha}\re^{-\CA/z} \sum_{k=0}^\infty b_k z^k,
\end{equation}
which is of the form of a non-perturbative trans-series.  It is
therefore clear that each singular point of the Borel transform, also
called a \emph{Borel singularity}, encodes a possible non-perturbative
contribution, which can be detected and extracted from the
perturbative series by calculating the Stokes discontinuity.
Alternatively, it can be shown from the formula for the Stokes
discontinuity that the large-order behavior of perturbative
coefficients $a_n$ is controlled by the non-perturbative
contribution through the so-called \emph{resurgence relation}
\cite{Bender:1990pd,dingle1973:ase}
\begin{equation}
  \label{eq:resurgence-relations}
  a_n  \sim %\sum_{i} \frac{\Gamma(n)}{\CA_i^n}\sum_{k}
            %\frac{b_k^{(i)}
    % \CA_i^{k}}{\prod_{j=1}^{k}(n-j)},
  \sum_i\sum_{k=0}^\infty\frac{\Gamma(n-\alpha_i-k)}{\CA_i^{n-\alpha_i-k}}b_k^{(i)},\quad n\rightarrow \infty,
\end{equation}
which can be used very efficiently to extract non-perturbative
information from the perturbative series.

One may wonder where these non-perturbative contributions come from.
It is helpful to recall the Lefschetz-thimble picture.  As the
orientation of the integration contour for the Borel resummation
changes, the associated Lefschetz thimble also changes its
orientation.  When the integration contour passes through a Borel
singularity, the Lefschetz thimble hits a neighboring saddle point,
after which it can flow in two different directions, corresponding to
the ambiguity of the two lateral Borel resummations.  The difference
between the two choices of Lefschetz thimble is exactly the Lefschetz
thimble attached to the neighboring saddle point, and therefore the
non-perturbative contributions discovered by the method of resurgence
in general correspond to neighboring saddle points in the exponential
integral \cite{berry:1991hyper}.%  \jjj{Distinction between finite
  % dimensional and infinite dimensional exponential integrals, and the
  % special case of one-dimensional integral. Recall the lecture by
  % Marcos Marino.}

There can be multiple Borel
singularities along a Stokes ray $d$.  The Stokes discontinuity across
the Stokes ray would receive non-perturbative contributions from all
these singular points.  This is encoded in the linear transformation
of trans-series called \emph{Stokes automorphism}
\begin{equation}
  \label{eq:stokes-automorphism}
  \mf{S}_d\varphi(z) = \varphi(z) + \sum_{\CA\in d}\re^{-\CA/z}\varphi^{(\CA)}(z)
\end{equation}
so that
\begin{equation}
  \mr{S}^{(\theta_+)}\varphi(z) = \mr{S}^{(\theta_-)}\mf{S}_d\varphi(z),
\end{equation}
and the Stokes discontinuity is given by
\begin{equation}
  \disc_d\varphi(z) = \mr{S}^{(\theta_-)}\left(\mf{S}_d-\md{1}\right)\varphi(z).
\end{equation}
To extract individual non-perturbative contributions associated to
each Borel singularity, one can define the so-called \emph{alien
  derivatives} associated to each singular point from the logarithm of
the Stokes automorphism \cite{Ecalle},
\begin{equation}\label{eq:logS-alien}
  \log \mf{S}_d = \sum_{\CA\in d}\dot{\Delta}_\CA.
\end{equation}
The alien derivative associated to a Borel singularity $\CA$ when
applied to the perturbative series serves to extract the particular
non-perturbative contribution with action $\CA$,
\begin{equation}
  \dot{\Delta}_\CA \varphi = \CS_\CA \re^{-\CA/z}
  \wt{\varphi}^{(\CA)}(z).
  \label{eq:alien}
\end{equation}
One can sometimes normalize the non-perturbative series
$\wt{\varphi}^{(\CA)}(z)$, following for instance the saddle point
expansion of the exponential integral or other considerations, and
there is then a proportionality coefficient in \eqref{eq:alien} which
is called the \emph{Stokes constant} $\CS_\CA$.  Note that the
Stokes automorphisms are group-like operations, while the alien
derivatives are Lie-algebra-like operations, and they satisfy various
interesting algebraic properties \cite{Ecalle}.  For instance, they
behave like derivations and satisfy both the Leibniz rule and the
chain rule.  This makes alien derivatives a powerful tool for
exploring non-perturbative contributions.

We have thus seen that perturbative series encode a rich set of
non-perturbative contributions, each of which is described by three
pieces of data: the action $\CA_*$, the non-perturbative series
$\varphi^{(*)}$, and the Stokes constant $\CS_*$.  Together they
form part of the \emph{resurgent structure} of the perturbative series
and can be extracted using the powerful toolkit of resurgence theory
including the Borel transform, resurgence relations, Stokes
automorphisms, and alien derivatives.  We can apply all these tools to
topological string free energy.  But let us first quickly review
perturbative topological string to set the stage.

\section{Perturbative topological string}
\label{sc:pt-TS}

A topological string theory model is defined on a family of Calabi--Yau
threefolds $X$ with varying K\"ahler moduli, and the perturbative free
energy is the generating series of Gromov--Witten invariants 
$N_{g,\md{d}}$
\begin{equation}
  \CF(\md{t},g_s) = \sum_{g=0}^\infty g_s^{2g-2}\CF_g(\md{t}),\quad
  \CF_{g}(\md{t}) = \sum_{\md{d}\in H_2(X,\IZ)} N_{g,\md{d}}\re^{-\md{d}\cdot\md{t}}.
\end{equation}
The Gromov--Witten (GW) invariants are in general rational numbers because of orbifold
actions on the moduli space of stable pseudoholomorphic maps. To obtain
integer invariants, one can use the Gopakumar--Vafa (GV) formula
\cite{Gopakumar:1998ii,Gopakumar:1998jq}
\begin{equation}
  \label{eq:GV}
  \CF(\md{t},g_s) = \sum_{g\geq 0}\sum_{\omega \geq 1}\sum_{\md{d}\in
    H_2(X)} \frac{n_{g,\md{d}}}{\omega}
  \left(2\sin\frac{\omega g_s}{2}\right)^{2g-2}
  \re^{-\omega \md{d}\cdot\md{t}},
\end{equation}
where the GV invariants $n_{g,\md{d}}$ are integers.

Alternatively, topological string theory can be defined on a family of mirror
Calabi--Yau threefolds $X^*$ with varying complex structure moduli.
The complex structure of a Calabi--Yau threefold is characterised by
period integrals,
\begin{equation}
  \Pi = (\CP_J,\CX^I)^T = \left(\int_{B^J} \Omega, \int_{A_I}\Omega \right)^T
\end{equation}
where $\Omega$ is the holomorphic $(3,0)$-form, and $A_I, B^J$ are a
symplectic basis of 3-cycles with the intersection numbers
\begin{equation}
  A_I \cap A_J = B^I\cap B^J = 0,\quad A_I\cap B^J = - B^J\cap A_I= \delta_{I}^J.
\end{equation}
The complex structure moduli space has singular points, around which
the period integrals undergo monodromy
\begin{equation}
  \Pi \rightarrow M\cdot \Pi,\quad M\in Sp(b_3,\IZ).
\end{equation}

There is a unique singular point called the maximally unipotent
monodromy point, or the MUM point, where the monodromy matrix has the
property that
\begin{equation}
  (M-\md{1})^{b_3} = 0,
\end{equation}
and we can choose a symplectic basis $A_I,B^J$ for the 3-cycles such
that $M$ is of the form
\begin{equation}
  M =
  \begin{pmatrix}
    1 & * & * & *\\
    0 & 1 & * & *\\
    0 & 0 & 1 & *\\
    0 & 0 & 0 & 1
  \end{pmatrix}
\end{equation}
Then $\{\CX_I\}$ can serve as a set of homogeneous flat
coordinates on the complex structure moduli space near the MUM point.
According to mirror symmetry, the MUM point of the complex structure
moduli space of $X^*$ corresponds to the large-volume limit of the
K\"ahler moduli space of $X$, and the period integrals are related to
the K\"ahler moduli by the mirror map
\begin{equation}
  t_i = \CX_i / \CX_0,\quad i = 1,\ldots,b_3-1.
\end{equation}

The perturbative topological string free energy can be calculated
rather efficiently.  Let us assume for simplicity that the moduli
space is of complex dimension one.  It turns out that for the free
energies $\CF_g(t)$ to be well-defined over the entire moduli space,
we have to promote them to almost holomorphic functions
$F_g(t,\bar{t})$\footnote{More precisely, sections of a line bundle
  over the moduli space.} which depend weakly on $\bar{t}$.  The
almost holomorphic free energies reduce to the holomorphic free
energies in the holomorphic limit
\begin{equation}
  \CF_g(t) = \lim_{\bar{t}\rightarrow\infty} F_g(t,\bar{t}).
\end{equation}
More importantly, the dependence on $\bar{t}$ is described by the
following partial differential equations known as the
Bershadsky--Cecotti--Ooguri--Vafa (BCOV) holomorphic anomaly equations
\cite{Bershadsky:1993ta,Bershadsky:1993cx}
\begin{equation}
  \pd_{\bar{t}}F_g = \frac{1}{2}C_{\bar{t}}^{tt}
  \left(D_t^2F_{g-1}+\sum_{h=1}^{g-1}D_tF_h D_t F_{g-h}\right),
  \label{eq:BCOV}
\end{equation}
where $C_{\bar{t}}^{tt}$ is the antiholomorphic Yukawa coupling, and
$D_t$ the covariant derivative.  The holomorphic anomaly equations
are recursive equations where the genera of the free energies that
appear on the right-hand side are strictly smaller than $g$.  We can
therefore solve for the free energies $F_g$ by direct integration
over $\bar{t}$ up to an integration constant known as the
holomorphic ambiguity that does not depend on $\bar{t}$.  The main
difficulty of solving holomorphic anomaly equations comes from fixing
the holomorphic ambiguity. Sometimes it can be completely resolved by,
for instance, considering the required asymptotic behavior of
$\CF_g(t)$ near the singular points of the moduli space, but in many
other examples, especially in compact models, these boundary
conditions are not enough.  There has been a lot of work put into
solving the holomorphic anomaly equations
\cite{Yamaguchi:2004bt,Huang:2006hq,Huang:2006si,Alim:2007qj,Grimm:2007tm,Haghighat:2008gw}.
It is now known that for topological string on local Calabi--Yau
threefolds such as local $\IP^2$ or local $\IF_0$, the holomorphic
ambiguities can in principle be fixed up to arbitrarily high genus
\cite{Haghighat:2008gw,Huang:2013yta}, limited only by the available computing time and disk space.  On a modern laptop, it
is possible to calculate the free energies up to genus 200 or beyond.
For models on compact Calabi--Yau threefolds such as the quintic, the
asymptotic behaviors of free energies near singular points only allow
us to determine $\CF_g(t)$ up to relatively low genus
\cite{Huang:2006hq}, and additional techniques are required.  Using
the correspondence with Donaldson--Thomas invariants, the free
energies of compact models such as the quintic have been calculated up
to around genus 60 \cite{Alexandrov:2023zjb,Alexandrov:2023ltz}.

%\jjj{Choice of frame?}

\section{Nonperturbative topological string}
\label{sc:np-TS}

\subsection{Conjecture}

% \jjj{Discuss that perturbative free energy of topological string is
%   1-Gevrey.}

As we discussed in the previous section, the perturbative
free energy of topological string theory on both local and compact Calabi--Yau threefolds
can be calculated to high orders.
It is therefore suitable to apply resurgence theory to extract
possible non-perturbative corrections.
Recall from section~\ref{sc:rs} that non-perturbative contributions
take the form of trans-series which include three pieces of
information,
\begin{itemize}
\item the non-perturbative action $\CA$ that determines the order of
  magnitude of the non-perturbative correction,
\item the non-perturbative series $\varphi^{(\CA)}$ that describes the
  fine details of the corrections,
\item and finally the Stokes constant $\CS_\CA$ that controls the
  Stokes automorphism relating the non-perturbative contribution to
  the perturbative series,
\end{itemize}
all of which can be extracted in principle from the perturbative
series using resurgence techniques.

\begin{figure}
  \centering%
  \subfloat[Local $\IP^2$]%
  {\includegraphics[height=4cm,valign=c]{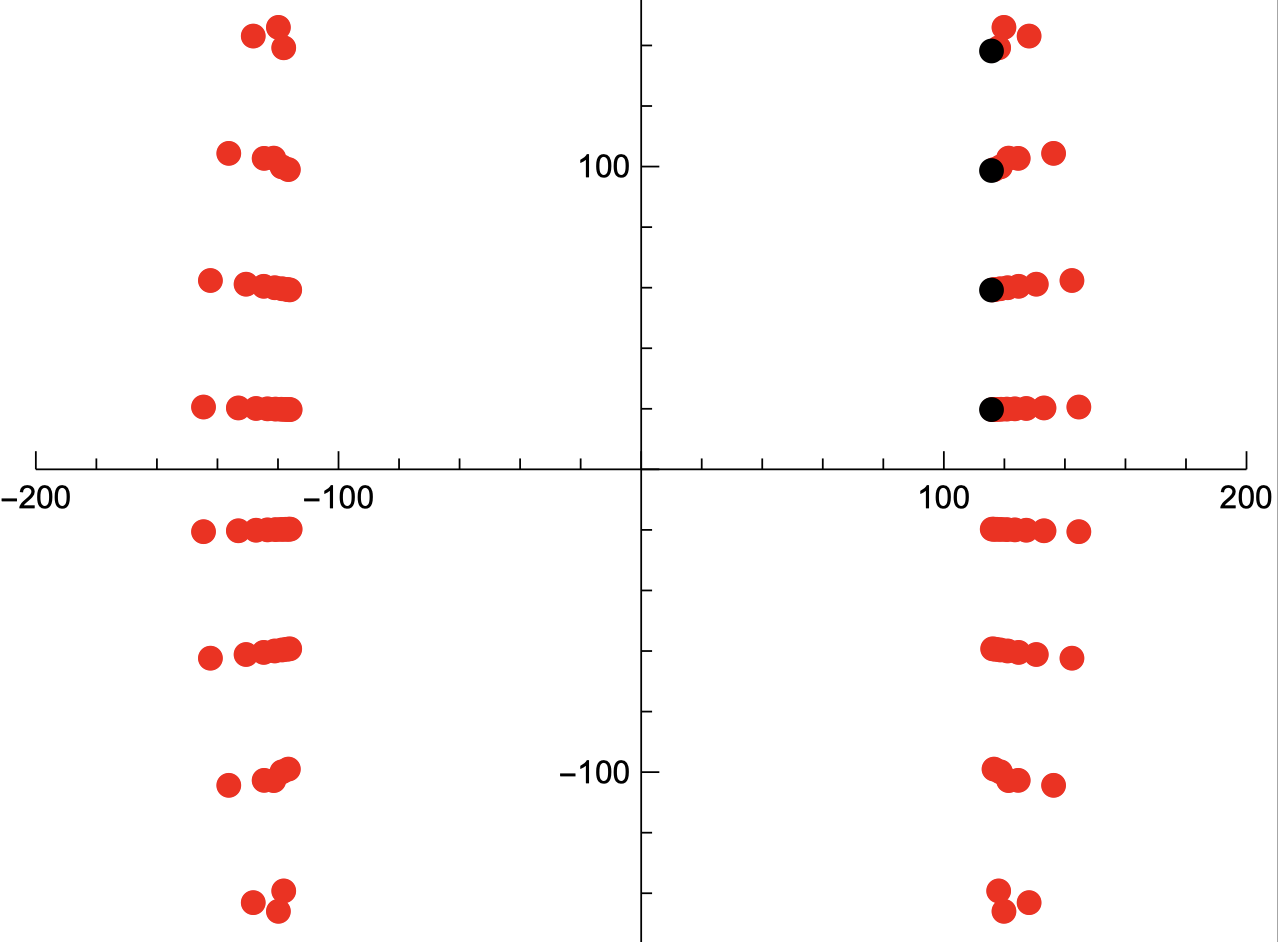}\label{fig:P2LR}}%
  \hspace{4ex}%
  \subfloat[Quintic]%
  {\includegraphics[height=4cm,valign=c]{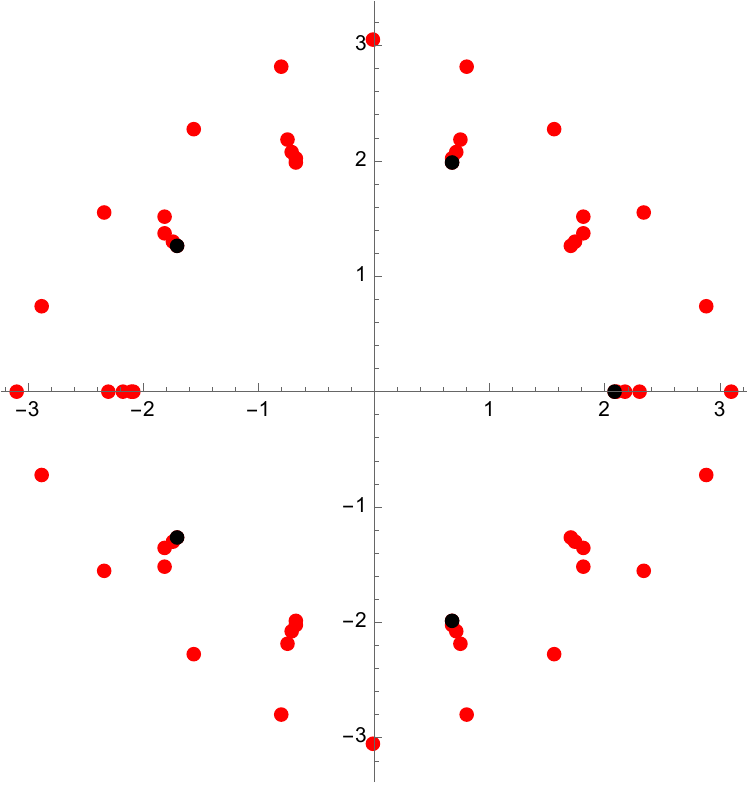}\label{fig:QntOrb}}%
  \caption{Borel singularities for (a) local $\IP^2$ near the Large
    Volume point and for (b) quintic near the orbifold point.  The red
    points are the singularities of the Pad\'e approximant which
    approximate branch cuts, and the black points mark the locations of the
    branch points.}
  \label{fig:CYbrl}
\end{figure}

The first thing one can do is to calculate the Borel transform of the
perturbative free energy and look for its singular points, i.e., the
Borel singularities, whose positions give the
non-perturbative actions of possible non-perturbative corrections.  As
we only have a finite number of perturbative coefficients,
approximation methods such as Pad\'e approximants are needed to
estimate the distribution of singularities.  This exercise can be done
for topological string models on both local and compact Calabi--Yau manifolds, and one notices immediately that the
non-perturbative action, after appropriate normalization, can always
be recognized as a linear combination of the integral basis of periods
with integer coefficients\footnote{It is important that when one moves
  in the moduli space without crossing a wall of marginal stability
  (see the discussion of stability below), the integer
  coefficients remain the same.},
\begin{equation}
  \CA_\gamma  = c^I \CP_I + d_I \CX^I,\quad \gamma =
  (c^I,d_I)\in\IZ^{b_3}.
  \label{eq:Agamma}
\end{equation}
For instance, the dominant Borel singularities for local $\IP^2$
near the large-volume point, and for the quintic near the orbifold
point are plotted in Fig.~\ref{fig:CYbrl}.  In the former case, they
are located at
\cite{Drukker:2011periods,Couso-Santamaria2016,Couso-Santamaria2015,Couso:2016quantization,Gu:2022sqc}
% \jjj{What normalization?
  % $\CX^1 =t ,\quad \CX^0 = 4\pi^2\ri$.}
\begin{equation}
  \CA = \pm \CX^1 + n \CX^0,\quad n =0,\pm1,\pm2,\ldots,
\end{equation}
while in the latter case, they are found to be \eqref{eq:Agamma} with
charges \cite{Gu:2023mgf}
\begin{equation}
  \gamma = \pm
  (0,0,0,1),\;\pm(0,0,-1,1),\;\pm(-1,-1,3,-1),\;\pm(2,1,-3,-2),\; \pm(-1,0,1,1).
\end{equation}
%
% \begin{Historyrevision}
% The period-integral proposal of \cite{Drukker:2011periods} and the
% local-model tests of
% \cite{Couso-Santamaria2016,Couso-Santamaria2015,Couso:2016quantization}
% preceded the closed-form multi-instanton results. The integral-period
% normalization and its extension to compact threefolds were developed
% in \cite{Gu:2023mgf}.
% \end{Historyrevision}

We thus have our first conjecture \footnote{There could be
  non-perturbative contributions of the order $\CO(\re^{-1/g_s^2})$
  which may come from Neveu--Schwarz branes, and whose action is the
  volume of the Calabi--Yau manifold.}:
\begin{coj}[\cite{Drukker:2011periods,Gu:2023mgf,Iwaki:2023rst}]
  The action $\CA$ of a non-perturbative contribution of the order
  $\CO(\re^{-\CA/g_s})$ to topological string free energy is an
  integral period.
\end{coj}

It is important to note that an integral period with charge vector
$\gamma$ can be identified with the central charge of D6-D4-D2-D0
brane bound states in type IIA superstring compactified on the same
Calabi--Yau threefold $X$\footnote{Or equivalently as D5-D3-D1-D(-1)
  brane bound states in the type IIB superstring compactified on the
  mirror Calabi--Yau threefold $X^*$.} where $c^0,c^i,d_i,d_0$ are
the charges of D6, D4, D2, and D0 branes, respectively.
This is the first indication that the non-perturbative contributions
to the topological string free energy are related to D-branes in
superstring theory.

The period description was first developed from the large-order
analysis of open strings and matrix models
\cite{Marino:2006hs,Marino:2007te,Marino:2008ya}.  The Borel
singularities and Stokes automorphisms of the conifold and
resolved-conifold free energies were determined in
\cite{Pasquetti2010}. Later, period integrals of the spectral curve
were proposed as spacetime instanton actions and tested in the matrix
model of ABJM theory \cite{Drukker:2011periods}. Subsequent
large-order studies of local $\IP^2$ provided tests beyond these
examples \cite{Couso-Santamaria2015,Couso:2016quantization}.  The
integral-period formulation for compact Calabi--Yau threefolds was
developed in \cite{Gu:2023mgf} and incorporated into the general
resurgent-structure conjecture in \cite{Iwaki:2023rst}; see also
\cite{Marino:2024tbx}.

Next, we consider the non-perturbative series. In principle, one could
use the resurgence relations to extract coefficients in the
non-perturbative series order by order from the perturbative series,
but this is far too laborious.  Instead, a more efficient method was
proposed in \cite{Couso-Santamaria2016,Couso-Santamaria2015}.  The
holomorphic anomaly equations \eqref{eq:BCOV} for the topological
string free energy can be recast in the master form
\begin{equation}
  \pd_{\bar{t}} F = \frac{g_s^2}{2} C_{\bar{t}}^{tt}
  \left(D_t^2F+ (D_tF)^2\right),
  \label{eq:BCOV-master}
\end{equation}
a partial differential equation whose solution is the
perturbative series
\begin{equation}
  F = F^{(0)} = \sum_{g\geq 1}g_s^{2g-2}F_g.
  \label{eq:Fpert}
\end{equation}
Consider moving in the complex $g_s$ plane.
When we cross a Stokes ray $d$ with for instance a sequence of
singular points $k\CA$ for $k=1,2,\ldots$,
the perturbative series will
undergo the Stokes automorphism $\mr{S}_d$ and become
\begin{equation}
  F^{(0)} \mapsto \mr{S}_d F^{(0)} =
  \exp\Big(\sum_{k=1}^\infty\dot{\Delta}_{k \CA}\Big)F^{(0)}
  = F^{(0)} +
  \dot{\Delta}_{\CA}F^{(0)} +\frac{1}{2}\dot{\Delta}^2_{\CA}F^{(0)} +
  \dot{\Delta}_{2\CA}F^{(0)} + \ldots,
  \label{eq:Fnp}
\end{equation}
and it is natural that the trans-series thus obtained should satisfy
the same differential equation.
Therefore, instead of \eqref{eq:Fpert} we can make the trans-series
ansatz for the solution to the holomorphic anomaly equation \eqref{eq:BCOV-master}
\begin{equation}
  F = F^{(0)} + F^{(1)} + F^{(2)} + \ldots
  \label{eq:F-ts}
\end{equation}
with
\begin{equation}
  F^{(0)} = \sum_{g\geq 1} g_s^{2g-2} F_g,\quad F^{(k)} =
  \re^{-k\CA/g_s}g_s^{-\nu_k}\sum_{n\geq 0} g_s^n F_n^{(k)},\;k\geq 1.
\end{equation}
and the alien derivatives $\dot{\Delta}_{k\CA}F^{(0)}$ can be read
off from \eqref{eq:F-ts} order by order in $\re^{-\CA/g_s}$.  For
instance,
% \jjj{Up to Stokes constants!}
\begin{align}
  F^{(1)} = \dot{\Delta}_{\CA}F^{(0)},\quad %
  F^{(2)} = \frac{1}{2}\dot{\Delta}^2_{\CA}F^{(0)} +
  \dot{\Delta}_{2\CA}F^{(0)},\quad\ldots.
  \label{eq:Fk-DelF0}
\end{align}
% \begin{subequation}
%   \begin{align}
%     &,\\
%     &,\\
%     &\ldots
%   \end{align}
% \end{subequation}

%

The non-perturbative series $F^{(k)}$ can be determined either order
by order in $g_s$ as done in
\cite{Couso-Santamaria2016,Couso-Santamaria2015} or in closed form by
exploiting the algebraic structure of the holomorphic anomaly
equations \cite{Gu:2022sqc,Gu:2023mgf}. % \footnote{There is an
  % issue of normalization of the non-perturbative series, which is the
  % non-perturbative version of the holomorphic ambiguity of the
  % perturbative series, and it is fixed using the resurgence relation.}
%
The results can be formulated rather elegantly in terms of the action of
the alien derivatives, from which the total non-perturbative
contribution can be reconstructed using \eqref{eq:logS-alien}. In terms of
the holomorphic partition function
\begin{equation}
  \CZ = \exp \CF,
\end{equation}
the alien derivative is given by \cite{Douaud:2026qfo}
\begin{equation}
    \label{eq:Znp}
    \dot{\Delta}_\gamma \CZ =
    \frac{\overline{\CS}_\gamma}{2\pi\ri}
    \big(1+\SD_\gamma\big)\re^{-\SD_\gamma}\CZ,
\end{equation}
where the operator $\SD_\gamma$ is
\begin{equation}
  \label{eq:SD}
  \SD_\gamma = g_s c^I \pd_{\CX^I} + \frac{1}{g_s}d_I\CX^I.
\end{equation}
Here the proportionality constant $\overline{\CS}_\gamma$ is the
Stokes constant, and it is the only unknown that cannot be fixed by
the holomorphic anomaly equations and must be determined via the
resurgence relation, as we discuss below.
The alien derivatives of the free energies follow from \eqref{eq:Znp}
immediately.  For the Borel singularities with $c^Ia_I\neq 0$ for any
$a_I\in \IC$, they take the form
\begin{equation}
  \dot{\Delta}_\gamma \CF = \frac{\overline{\CS}_\gamma}{2\pi\ri}
  \left(1+g_s c^I\pd_{\CX^I}\CF^\#(\CX^I-g_sc^I)\right)\exp(\CF^\#(\CX^I-g_sc^I)-\CF^{\#}(\CX^I)),
\end{equation}
where $\CF^{\#}$ is the modified free energy
\begin{equation}
  \CF^{\#}(\CX^I) = \CF(\CX^I) +
  \begin{cases}
    \frac{1}{2g_s^2 c^Id_I}(d_I\CX^I)^2,\quad \text{if }c^Id_I\neq
    0,\\
    \frac{1}{g_s^2c^Ia_I}(a_I\CX^I)(d_J\CX^J),\quad \text{for any }a_I:
    c^Ia_I\neq 0 \;\text{ if } c^Id_I = 0.
  \end{cases}
\end{equation}
On the other hand, for the Borel singularities with $c^I = 0$, the
alien derivative is particularly simple and
truncates to a two-term polynomial,
\begin{equation}
  \dot{\Delta}_\gamma\CF =
  \frac{\overline{\CS}_\gamma}{2\pi\ri}\Big(1+\frac{\CA_\gamma}{g_s}\Big)\re^{-\CA_\gamma/g_s}.
  \label{eq:F-PS}
\end{equation}
This expression was first discovered in \cite{Pasquetti2010} and is known as
the Pasquetti--Schiappa form.
Note that the total non-perturbative contributions \eqref{eq:Fnp}
involve repeated action of the alien derivatives, which can be
calculated by exploiting the analytic properties of the alien
derivatives as derivations.

Strictly speaking, it is not guaranteed that the non-perturbative
contributions obtained from the trans-series solution \eqref{eq:F-ts}
to the holomorphic anomaly equations are in the resurgent structure of
the perturbative free energy, as the resurgence theory only captures
\emph{some} non-perturbative contributions which are necessary to make
the Borel resummation of the perturbative series consistent, and there
can be additional non-perturbative contributions which are invisible
to resurgence theory applied to the perturbative
series. \footnote{A typical example is the 1-instanton correction
  to the energy series of the double-well quantum mechanical model.}
The statement that the non-perturbative contributions to the topological
string free energy in the same resurgent structure as
the perturbative free energy can be found as solutions to the
trans-series form of the holomorphic anomaly equations is therefore a
conjecture and must be checked explicitly through either the
resurgence relations \eqref{eq:resurgence-relations} or the Stokes
discontinuities \eqref{eq:stokes-automorphism}.

\begin{coj}[\cite{Couso-Santamaria2016,Couso-Santamaria2015,Gu:2022sqc,Gu:2023mgf}]
  The power series in non-perturbative contributions to topological
  string free energy which are related to the perturbative free energy
  by Stokes transformations are trans-series solutions to the
  holomorphic anomaly equations.
\end{coj}

% Resurgent trans-series techniques were first developed in
% \cite{Aniceto2012}.  The trans-series extension of the holomorphic
% anomaly equations was introduced in \cite{Couso-Santamaria2016}, and
% its resurgent interpretation was tested in local $\IP^2$ in
% \cite{Couso-Santamaria2015}. Comparison with spectral theory
% subsequently tested the resummed trans-series
% \cite{Couso:2016quantization}. Closed-form multi-instanton amplitudes
% were obtained for local threefolds in \cite{Gu:2022sqc} and extended
% to compact threefolds in \cite{Gu:2023mgf}, while the resulting Stokes
% automorphism and its action on the dual partition function were worked
% out in \cite{Iwaki:2023rst}.

This conjecture has passed many non-trivial tests both in
noncompact models \cite{Gu:2022sqc} and compact models
\cite{Gu:2023mgf} of topological string theory.  Along the way, it was
found that with an appropriate normalization of the non-perturbative
amplitudes, as indicated by \eqref{eq:Znp}, the Stokes constants
$\overline{\CS}_\gamma$ have an interesting integrality structure and
a natural physical interpretation.  Let us define the
primitive Stokes constants $\CS_\gamma$ via
\begin{equation}
  \overline{\CS}_\gamma = \sum_{n|\gamma} \frac{\CS_{\gamma/n}}{n^2}.
  \label{eq:multcover}
\end{equation}
Then we have the following conjecture.

\begin{coj}[\cite{Gu:2021ize,Iwaki:2023rst,Marino:2024yme}]
  The primitive Stokes constants of topological string free energy are
  BPS invariants or generalized Donaldson--Thomas invariants
  \begin{equation}
    \CS_\gamma = \Omega(\gamma).
  \end{equation}
\end{coj}

The BPS invariants or the generalized DT invariants are the
multiplicities of stable D6-D4-D2-D0 bound states in type IIA
superstring theory.  Some typical examples are summarized in
Tab.~\ref{tab:DT}.  This also answers the question of what D-brane
bound states can appear as Borel singularities: they should be stable
D-brane bound states.

\begin{table}
  \centering
  \begin{tabular}{cc}\toprule
    $\gamma$ & $\Omega(\gamma)$\\\midrule
    -1D6-D2-D0 & Pandharipande--Thomas (PT) invariants \cite{Pandharipande:2007PT}\\
    1D6-D2-D0 & DT invariants \cite{Thomas:1998DT,Maulik:2003GWDT}\\
    D2-D0 & GV invariants $n_{0,\md{d}}$ \cite{Gopakumar:1998ii,Gopakumar:1998jq}\\
    D4-D2-D0 & Maldacena--Strominger--Witten (MSW) invariants \cite{Maldacena:1997MSW}\\\bottomrule
  \end{tabular}
  \caption{Generalized DT invariants}
  \label{tab:DT}
\end{table}

The connection with BPS states was anticipated in Mari\~no's Strings
2019 talk \cite{Marino:Strings2019}. Integer Stokes data in complex
Chern--Simons theory
\cite{Garoufalidis:2020knots,Garoufalidis:2020peacock,Garoufalidis:2021trivial}
motivated the proposal in \cite{Gu:2021ize} of integer Stokes
invariants for a special conifold expansion of topological strings.
Explicit resolved-conifold comparisons followed in
\cite{Alim:2021mhp,Grassi:2022zuk,Alim:2022curves}.
%
% General
% formulations of the proposed identification appear in
% \cite{Iwaki:2023rst,Marino:2024yme}.  Direct calculations for compact
% models, wall-crossing, modular properties, and Wilson-loop insertions
% provide further evidence
% \cite{Douaud:2024khu,Douaud:2026qfo,Guo:2026xpg,Gu:2024wag}.
%
% Substantial evidence has accumulated for this conjecture. The conifold
% and resolved-conifold calculations of \cite{Pasquetti2010} supply
% basic examples. For the resolved conifold, the relation of Stokes
% jumps to DT invariants was made explicit in \cite{Alim:2021mhp}.
%
Another piece of evidence comes from the large-radius analysis of the
GV representation \eqref{eq:GV}
\cite{Pasquetti2010,Aniceto2012,Couso:2016quantization,Gu:2023mgf}.
Near the large-volume point and in the large-volume frame, this
analysis motivates the following genus-zero-GV contribution to the
large-genus asymptotics:
\begin{equation}
  \CF_g(\CX) \sim \sum_{\md{d}\in H_2(X)}\sum_{d_0\in\IZ}
  \frac{n_{0,\md{d}}}{2\pi^2}\frac{(2g)!}{(d_i\CX^i
    +d_0\CX^0)^{2g-2}}\frac{1}{2g(2g-2)} +\ldots
\end{equation}
Comparison with the resurgence relation
\eqref{eq:resurgence-relations} suggests, in this regime,
singularities associated with D2-D0 states of charge
$\gamma=(0;\md{d},d_0)$, at
\begin{equation}
  \CA_\gamma = d_i\CX^i + d_0 \CX^0,
\end{equation}
and the associated non-perturbative series is of the
Pasquetti--Schiappa form \eqref{eq:F-PS}, while the primitive Stokes
constant is the GV invariant
\begin{equation}
  \CS_\gamma = n_{0,\md{d}} = \Omega(0;\md{d},d_0).
\end{equation}
This is also illustrated in the Borel plane plot in
Fig.~\ref{fig:P2LR} of local $\IP^2$ near the large-volume point,
where the two towers of Borel singularities correspond to the BPS
states with brane charges $\pm(0;1,n)$ with $n\in\IZ$.

\begin{figure}
  \centering
  \subfloat[$\CS_{(1;0,0)}=1$]{\includegraphics[height=4cm]{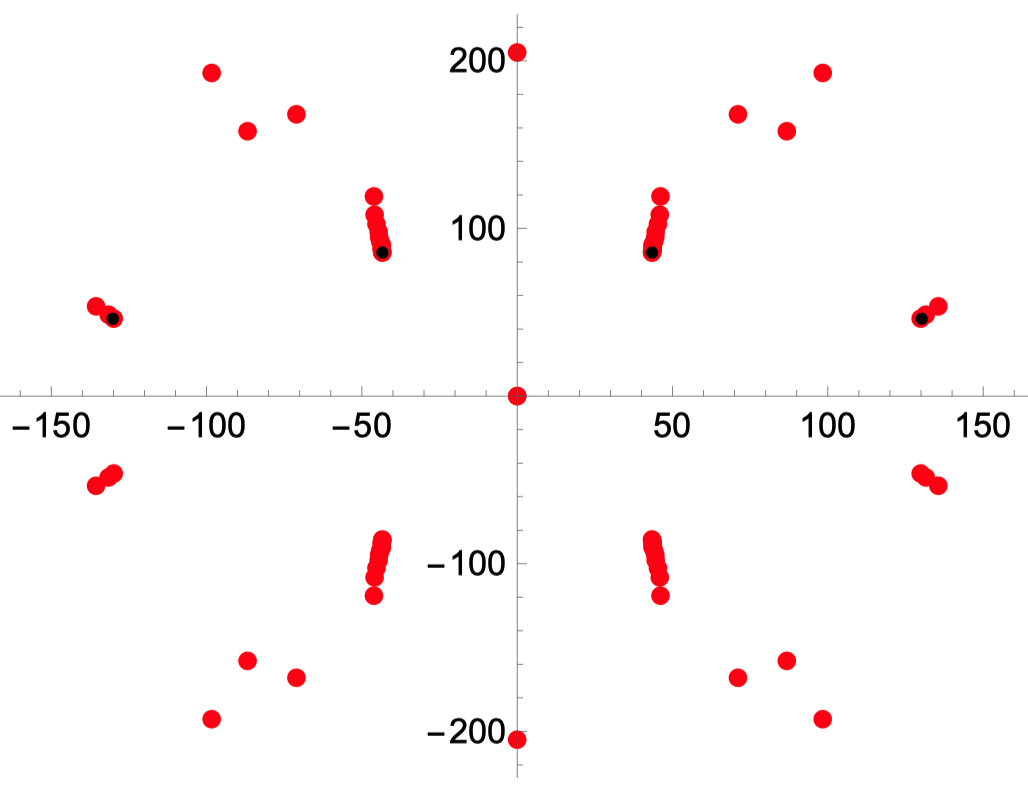}
  \label{fig:P2brl0}}\hspace{4ex}
  \subfloat[$\CS_{(1;0,1)}=3$]{\includegraphics[height=4cm]{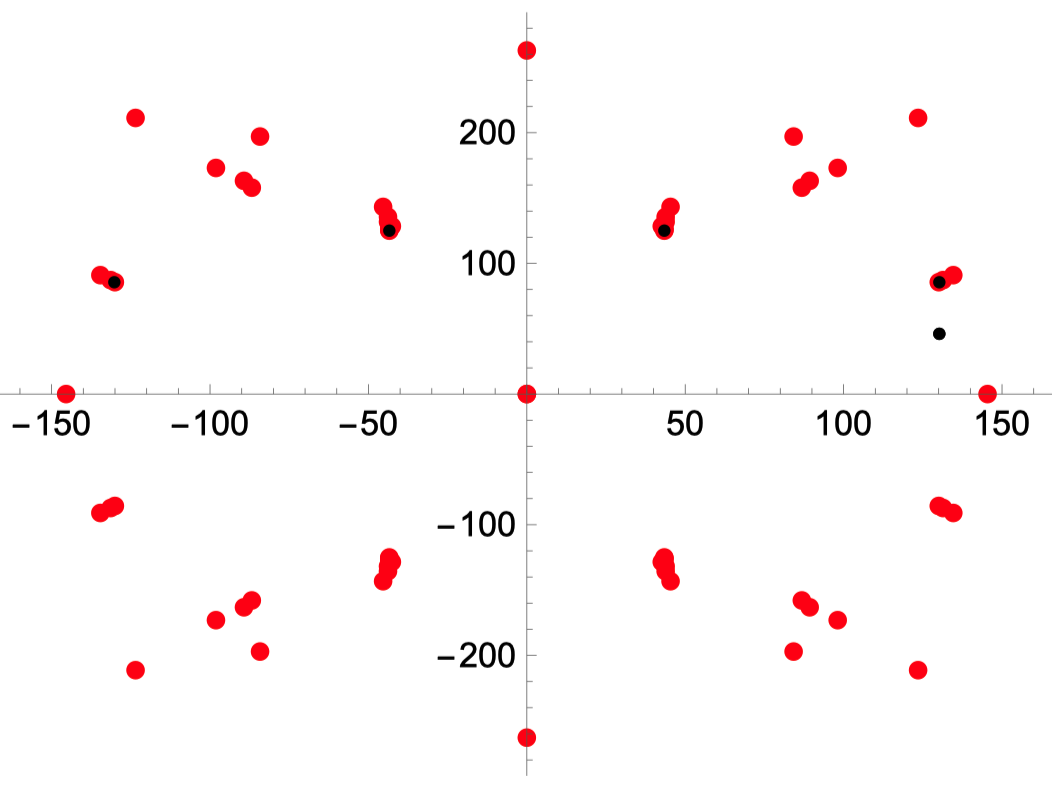}}\\
  \subfloat[$\CS_{(1;0,2)}=9$]{\includegraphics[height=4cm]{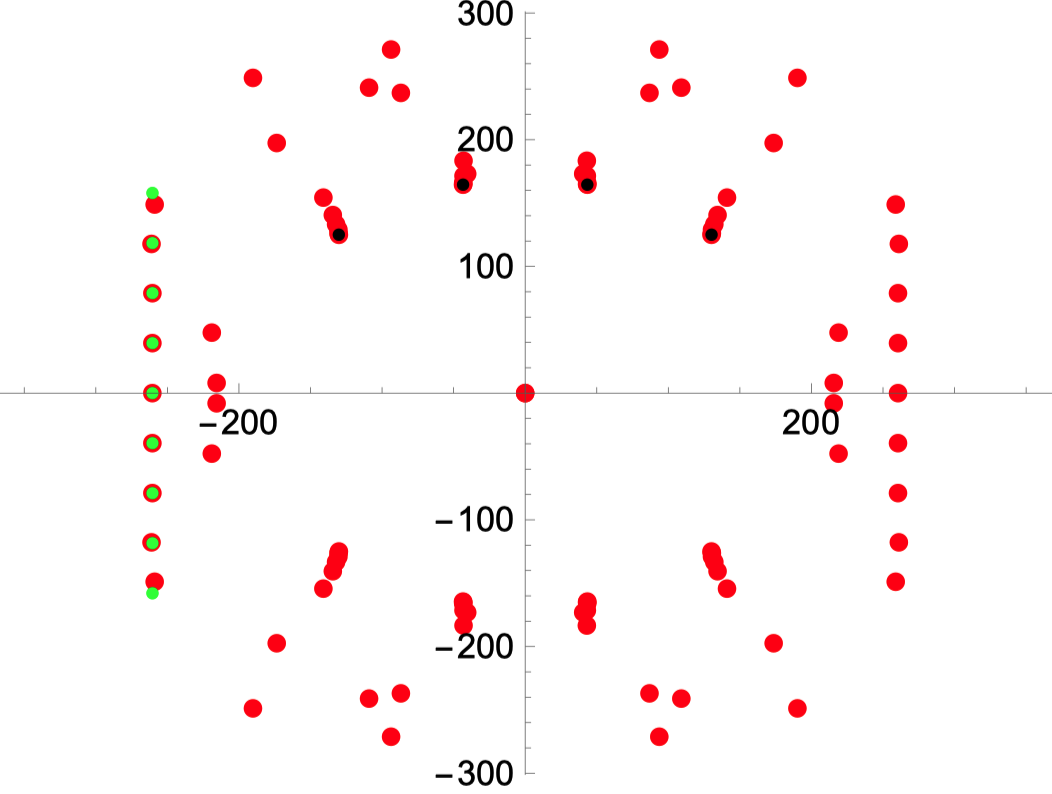}}\hspace{4ex}
  \subfloat[$\CS_{(1;0,3)}=22$]{\includegraphics[height=4cm]{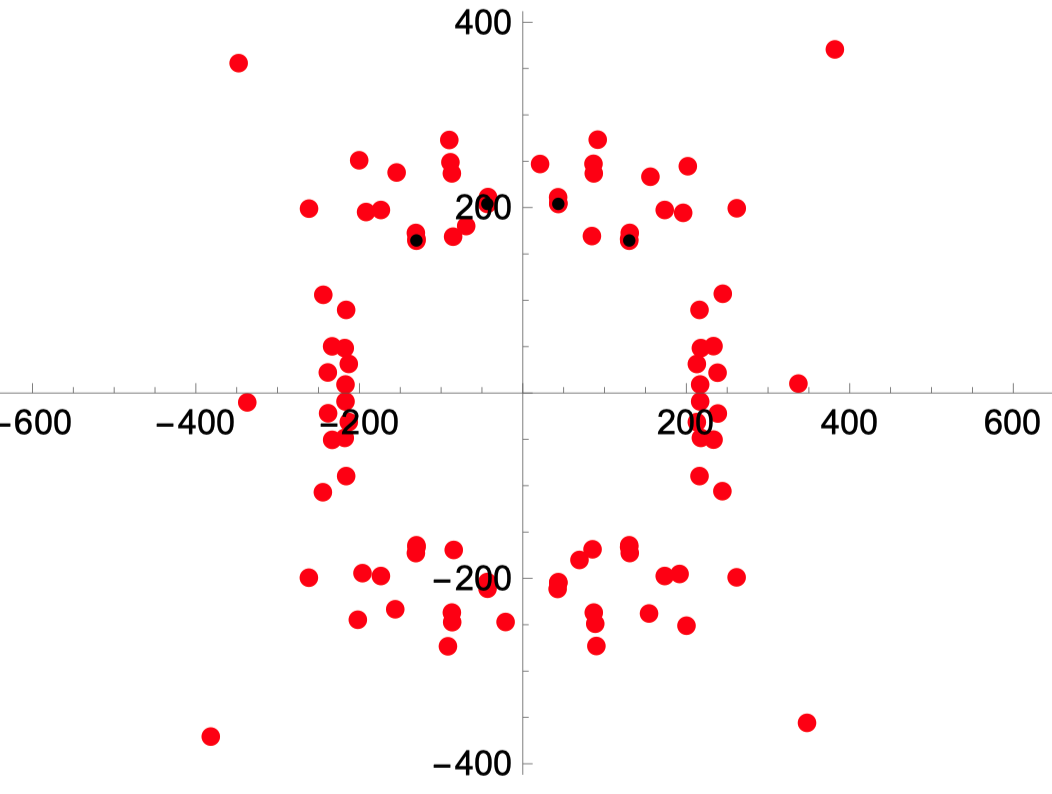}}
  \caption{Borel singularities of local $\IP^2$ near the large-volume
    point (with $z>0$) after successive subtraction of singularities.}
  \label{fig:P2brl}
\end{figure}

%\begin{Historyrevision}
The conifold locus gives complementary evidence. The universal
singular behavior of the free energies is related to the non-critical
$c=1$ string \cite{Ghoshal:1995conifold}, while its physical
interpretation involves a BPS hypermultiplet becoming massless
\cite{Strominger:1995conifold,Vafa:1995conifold}. In the normalization
of \cite{Pasquetti2010}, the associated primitive Stokes constant is
one, consistent with the index of that
hypermultiplet. % This is a local
% conifold statement, not a determination of the full BPS spectrum.
% \end{Historyrevision}

Further non-trivial numerical evidence has been found.  For instance,
the contributions of the non-perturbative sectors $(0;\md{d},d_0)$
in the resurgence relations can be removed, and one can find
additional Borel singularities.  For instance, in the case of local
$\IP^2$ near the large-volume point, one finds eight singular points
as in Fig.~\ref{fig:P2brl0}, which are related to each other by
monodromy action, as we will explain in the next subsection, and they
share the same Stokes constant 1.  One of the Borel singularities has
brane charge $\gamma = (1;0,0)$ and we thus have
\cite{Douaud:2026qfo}
\begin{equation}
  \CS_{(1;0,0)} = 1,
\end{equation}
which agrees with the DT invariant of this BPS state
\cite{Bousseau:2022snm}
\begin{equation}
  \Omega(1;0,0) = 1.
\end{equation}
After subtracting the contributions of these eight
non-perturbative sectors, eight further singularities appear in the
Borel plane, which also share the same Stokes constant.  One of the
new singularities carries brane charge $\gamma = (1;0,1)$ and we find
\begin{equation}
  \CS_{(1;0,1)} = 3,
\end{equation}
which also agrees with the DT invariant with the same brane charge.
The same exercise can be repeated two more times, yielding new Stokes
constants that agree with DT invariants, until numerical precision is
lost \cite{Douaud:2026qfo}.  Similar agreement has been found in the
topological string model of $\mathbb{F}_0$ \cite{Douaud:2026wip} and
in compact models \cite{Douaud:2024khu}.

Calculations of the Stokes constants have also been performed
elsewhere in moduli space, for instance near the conifold point.  See,
for instance, the discussions in
\cite{Gu:2022sqc,Gu:2023mgf,Marino:2023gxy,Marino:2024yme,Douaud:2024khu,Douaud:2026qfo}.

% \jjj{Results of TS/ST correspondence? Wall-crossing at conifold
%   point?}

% \subsection{Evidences}

\subsection{Modular properties}
\label{sc:modularity}

Except for simple models such as the resolved conifold, we only have a
finite number of perturbative free energies, and thus numerical
calculations of the Borel transform can reveal only a finite number
of non-perturbative sectors, while the number of BPS states is usually
infinite.  To facilitate the comparison between the two sides, it is
helpful to exploit the modular properties of non-perturbative
contributions to topological string theory, which can sometimes generate
an infinite number of Stokes constants.

Let us state what we mean by modularity.  As we mentioned before, the
moduli space of a topological string model typically has singular
points, and integral periods undergo monodromy action $M$ when we move
around a singular point and return to the starting point.  All
the monodromy actions together generate the monodromy group $\Gamma$
which is a subgroup of $Sp(b_3,\IZ)$.

The moduli space has additional structure. It is
criss-crossed by walls of real codimension one of marginal stability and
divided by them into different stability chambers. The BPS spectrum is
the same within a stability chamber but changes across a wall of
marginal stability.  Furthermore, if a stability chamber has singular
points inside, we say that their monodromy actions generate a
\emph{local} monodromy group.  When we move around such an internal
singular point, the associated monodromy action will permute the basis
of integral periods and thus change the brane charges of stable brane
bound states.  Therefore the stable brane bound states must form
orbits of the local monodromy group, and the bound states within each
orbit share the same BPS invariant.

\begin{figure}
  \centering%
  \subfloat[Moduli space]%
  {\includegraphics[height=4cm]{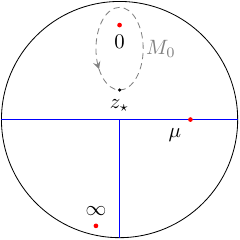}}%
  \hspace{4ex}
  \subfloat[Pert.\
  and non-pert.\ sectors]%
  {\includegraphics[height=4cm]{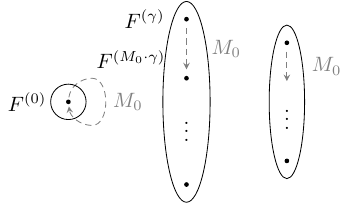}}%
  \caption{Modular covariance of non-perturbative contributions.}
  \label{fig:modularity}
\end{figure}

Let us now consider the implications of these structures in moduli
space for the perturbative free energy of topological string theory and
its non-perturbative contributions.  For simplicity, we restrict ourselves
to local Calabi--Yau threefolds.  It is known that the
perturbative free energy is invariant under actions of the global
monodromy group \cite{Bershadsky:1993cx,Aganagic:2006wq}. What about
non-perturbative contributions?  They are in general not invariant,
but it was shown that they transform covariantly under actions of
local monodromy groups \cite{Guo:2026xpg}.  This means that when we
move around a singular point inside a stability chamber with the local
monodromy action $M$, which is a $b_3\times b_3$ integer-valued
matrix, the non-perturbative sector labeled by the brane charge vector
$\gamma$ is transformed to the sector labeled by the brane charge
vector
\begin{equation}
  \gamma \mapsto \gamma \cdot M,
\end{equation}
and the non-perturbative amplitude transforms to
\begin{equation}
  F^{(\gamma)} \mapsto F^{(\gamma\cdot M)},
\end{equation}
which is precisely the non-perturbative amplitude in the
non-perturbative sector after monodromy transformation.  This is
illustrated in Fig.~\ref{fig:modularity}.  More importantly, it can be
shown that the Stokes constant remains the same, i.e.,
\begin{equation}
  \CS_\gamma \mapsto \CS_{\gamma\cdot M} = \CS_\gamma.
\end{equation}
Therefore, within each stability chamber, the non-perturbative
sectors are grouped into orbits of the local monodromy group, and
non-perturbative contributions in the same orbit share the same Stokes
constant.

These transformation properties of non-perturbative sectors under
local monodromy groups have two important implications.  First, some
orbits of local monodromy groups may have infinitely many
non-perturbative sectors and we are thus able to deduce the
Stokes constants associated to all of them even if we initially know only a few.  Secondly, it is clear that the transformation properties of the
non-perturbative contributions and the associated Stokes constants
under the local monodromy group inside a stability chamber
match those of the stable brane bound states and the associated BPS
invariants, and it is natural then to identify the two sides.

\begin{figure}
  \centering%
  \subfloat[Moduli
  space]{\includegraphics[height=3.3cm]{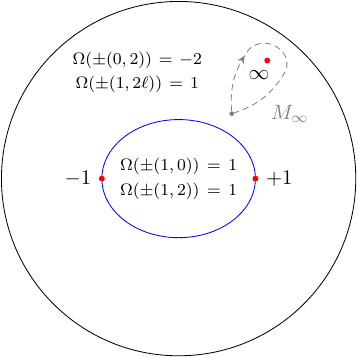}\label{fig:SWCB}}%
  \hspace{0.35cm}%
  \subfloat[Borel plane\$strong coupling)]{\includegraphics[height=3.3cm]{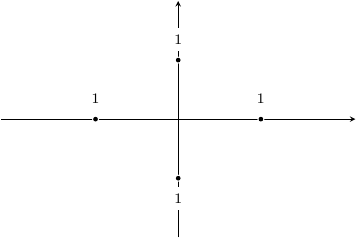}\label{fig:SWStokesStrong}}%
  \hspace{0.35cm}%
  \subfloat[Borel plane\$weak coupling)]{\includegraphics[height=3.3cm]{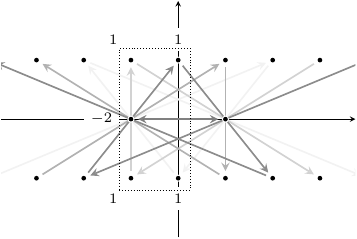}\label{fig:SWStokes}}%
  \caption{BPS invariants and resurgent structure of Seiberg--Witten
    theory. The Borel planes are schematic; the numerical labels in
    (b) and (c) indicate Stokes constants.}
  \label{fig:SW}
\end{figure}

Let us look at some examples.  A standard example is four-dimensional (4d)
$\CN=2$ Seiberg--Witten theory.  Its Coulomb branch moduli space -- the
$u$ plane -- has three singular points $u=\pm\Lambda$ and $u=\infty$,
whose monodromies generate the monodromy group $\Gamma = \Gamma(2)$.
It has a wall of marginal stability that passes through the two
singular points $\pm\Lambda$ and the wall divides the moduli space into two stability
chambers as shown in Fig.~\ref{fig:SWCB}.

Inside the wall of marginal stability is the strong-coupling chamber
with small $u$, where the BPS
states include two hypermultiplets and their charge-conjugation, parity, and time-reversal (CPT) conjugates, which contain
respectively the magnetic monopole of charge $(1,0)$ and the dyon of
charge $(1,2)$.  The BPS spectrum is then
\begin{equation}
  \Omega(\pm(1,0)) = 1,\quad \Omega(\pm(1,2)) = 1.
  \label{eq:SWBPS-smallu}
\end{equation}
These BPS states transform trivially under the local monodromy group
which contains only the identity element as there is no singular point
inside the strong-coupling chamber.

Outside the wall of marginal stability is the weak-coupling chamber
with large $u$, where the BPS states include a vector multiplet and
its CPT conjugate that have $W^{\pm}$ bosons of charges $\pm (0,2)$ as well
as infinitely many hypermultiplets that contain dyons of unit magnetic
charge and various electric charges $\pm (1,2\ell)$ for $\ell\in\IZ$.
The BPS spectrum reads
\begin{equation}
  \Omega(\pm(0,2)) = -2,\quad \Omega(\pm(1,2\ell)) = 1,\;\ell\in\IZ.
  \label{eq:SWBPS-largeu}
\end{equation}
On the other hand, the weak-coupling chamber does contain the singular
point $u=\infty$, whose monodromy action is represented by the
monodromy matrix
\begin{equation}
  M_\infty =
  \begin{pmatrix}
    -1 & 4 \\ 0 & -1
  \end{pmatrix},
  \label{eq:SWM}
\end{equation}
and it generates the non-trivial local monodromy group
$\Gamma_\infty = \IZ$, which is a subgroup of $\Gamma$.  It can be
shown that the pair of vector multiplets form a doublet of
$\Gamma_\infty$, and the hypermultiplets form four infinite orbits
with charges
\begin{subequations}
  \begin{align}
    &\{(-1,0),\;(1,-4),\;(-1,8),\;\ldots\},\\
    &\{(1,0),\;(-1,4),\;(1,-8),\;\ldots\},\\
    &\{(-1,-2),\;(1,-2),\; (-1,6),\;\ldots\},\\
    &\{(1,2),\;(-1,2),\;(1,-6),\;\ldots\}.
  \end{align}
\end{subequations}

This BPS spectrum can be reproduced exactly from the resurgent
structure of the topological string model defined on the
Seiberg--Witten curve.  The perturbative free energy can be calculated
very efficiently by solving the holomorphic anomaly equations
\cite{Huang:2006si}, and its Borel plane can be studied as follows
\cite{Marino:2024yme,Guo:2026xpg}.

For small values of $u$, the Borel plane displays four
singular points in the neighborhood of the origin, as illustrated in
Fig.~\ref{fig:SWStokesStrong}. When decomposed
in terms of the basis of integral periods their coefficients agree
precisely with the charges of the four BPS states in
\eqref{eq:SWBPS-smallu}, and in addition the Stokes constants coincide
with the BPS invariants.

For large values of $u$, the Borel plane displays many more singular
points, as shown in Fig.~\ref{fig:SWStokes}.  Their
charges agree with those of the BPS states in
\eqref{eq:SWBPS-largeu}, and they are related by the local
monodromy action \eqref{eq:SWM} according to Fig.~\ref{fig:SWStokes},
so that the Stokes constants of the five Borel singularities in
the dotted box, which have brane charges
\begin{equation}
  \gamma = (0,-2),\; (\pm 1,0),\; (\pm 1,-2),
\end{equation}
determine the entire spectrum of Stokes constants, which again
coincide with the BPS invariants.

\begin{figure}
  \centering%
  \subfloat[Moduli space]{\includegraphics[height=3.5cm]%
    {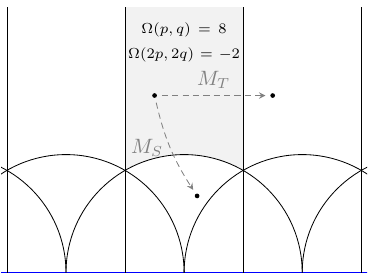}\label{fig:Nf4CB}}%
  \hspace{6ex}%
  \subfloat[Borel plane]{\includegraphics[height=4cm]%
    {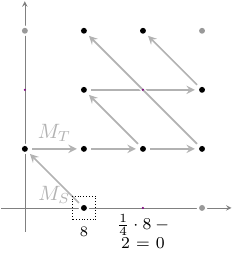}\label{fig:Nf4Stokes}}%
  \caption{BPS invariants and resurgent structure of 4d $\CN=2$
    $SU(2)$ massless $N_f=4$ supersymmetric quantum chromodynamics (SQCD).}
  \label{fig:Nf4}
\end{figure}

As a second example, we consider 4d $\CN=2$ SQCD with gauge group
$SU(2)$ and four massless hypermultiplets in the fundamental
representation, the $N_f= 4$ theory.  This is a superconformal theory,
and its complexified gauge coupling $\tau$ does not change under the
renormalization group (RG) flow.  The BPS spectrum of this theory consists of the following
\cite{Ferrari:1997gu}:
\begin{itemize}
\item Hypermultiplets of charges $(p,q)$ where $p,q$ are coprime
  integers. They include fundamental quarks of charge $\pm(0,1)$ which
  transform as $\md{8}_v$ of the flavor symmetry group $SO(8)$, and
  monopoles of charge $\pm(1,0)$ which transform as either $\md{8}_s$
  or $\md{8}_c$ of $SO(8)$. Each hypermultiplet contributes one to the
  BPS invariant, and thus the total BPS invariant for each charge
  $(p,q)$ is
  \begin{equation}
    \Omega(p,q) = 8.
  \end{equation}
\item Vector multiplets of charges $(2p,2q)$ where $p,q$ are coprime
  integers. They include the $W^{\pm}$-bosons of charge $\pm (0,2)$,
  which transform as singlets of the flavor symmetry group. Each
  vector multiplet contributes $-2$ to the BPS invariant, and thus for
  each charge $(2p,2q)$, the BPS invariant is
  \begin{equation}
    \Omega(2p,2q) = -2.
  \end{equation}
\end{itemize}
The full BPS spectrum is in fact manifestly self-dual, invariant under
$SL(2,\IZ)$ transformations. Conversely, assuming $SL(2,\IZ)$
invariance, the entire BPS spectrum can be generated by two states,
the hypermultiplet of charge $\pm(1,0)$ and the vector multiplet of
charge $\pm(0,2)$. This is illustrated in Fig.~\ref{fig:Nf4CB}.

Likewise, we can consider the topological string model associated with
SQCD, compute the perturbative free energy using the holomorphic
anomaly equations \cite{Huang:2011qx}, and study its resurgent
structure.  The results are as follows \cite{Guo:2026xpg}.  The Borel
singularities of the free energy are plotted in
Fig.~\ref{fig:Nf4Stokes}.  They are related to each other by
$SL(2,\IZ)$ transformations, and from the Stokes constant of the
single Borel singularity in the dotted box on the positive real axis,
which is 8, we can infer the Stokes constants of all these Borel
singularities in the same monodromy orbit, which agree with the BPS
invariants of the hypermultiplets.  Furthermore, the 2-instanton sectors
of these Borel singularities are absent, for instance at the second point
on the positive real axis, which indicates that there is a
1-instanton sector of a new Borel singularity, whose primitive Stokes
constant is such that it cancels the one corresponding to the
hypermultiplet.  Using the multi-covering formula
\eqref{eq:multcover}, one concludes that the primitive Stokes constant
of these new Borel singularities should be $-2$, which coincides with
the BPS invariant of the vector multiplet; these singularities also have
the correct charges.  Therefore, the Borel singularities together with
the associated Stokes constants again reproduce the entire BPS
spectrum of the supersymmetric theory.

\subsection{Supporting arguments}
\label{sc:arguments}

The three conjectures give a coherent picture which suggests a
correspondence between the non-perturbative contributions to
topological string free energy and stable D-brane bound states, i.e.,
BPS states, in type II superstring theory:
\begin{itemize}
\item the action of a non-perturbative contribution is the central
  charge of the BPS state;
\item the Stokes constant is the multiplicity of the BPS state.
\end{itemize}
But how do we understand this correspondence?  Several
approaches have been proposed, some of which we discuss below.  We will be
rather schematic, and refer to the cited references for details.

\subsubsection{Exact WKB and the Nekrasov--Shatashvili (NS) limit}

\begin{figure}
  \centering
  \begin{tikzpicture}
    \node (BPS) at (0,0) [align=center] {\textbf{BPS invariants
        $\Omega(\gamma)$}\\of Calabi--Yau threefold};%
    \node (QP) [right=3 of BPS,align=center] {Stokes coefficients
      $\mc{S}_{\gamma}^{\text{QP}}$ \\of quantum periods};%
    \node (NS) [below=1 of QP,align=center] {Stokes coefficients
      $\mc{S}_{\gamma}^{\text{NS}}$\\of NS topological string};%
    \node (TS) [below=1.05 of BPS,align=center] {\textbf{Stokes
        coefficients $\mc{S}_{\gamma}^{\text{top}}$}\\of conv'l
      topological string};%
    \draw [<->] (TS) -- node [below]{\small blowup eq'n} (NS);%
    \draw [<->] (NS) -- node [right]{\small qu'n special geometry}
    (QP);%
    \draw [<->] (QP) -- node [above]{\small geodesic cycles} (BPS);%
    \draw [<->,densely dashed] (BPS) -- (TS);%
    % \node (qp) at (0,0) {Stokes coefficients of quantum
    % periods}; % \node [right = 0.2 of qp] {$=$}; % \node (BPD) [right=of qp] {BPS invariants}; % \node [below = 0.2 of qp] {$||$}; % \node (NS) [below=of qp] {Stokes coefficients of NS free energies}; %
  \end{tikzpicture}
  \caption{Identification of Stokes constants and BPS invariants via
    NS limit.}
  \label{fig:NS}
\end{figure}
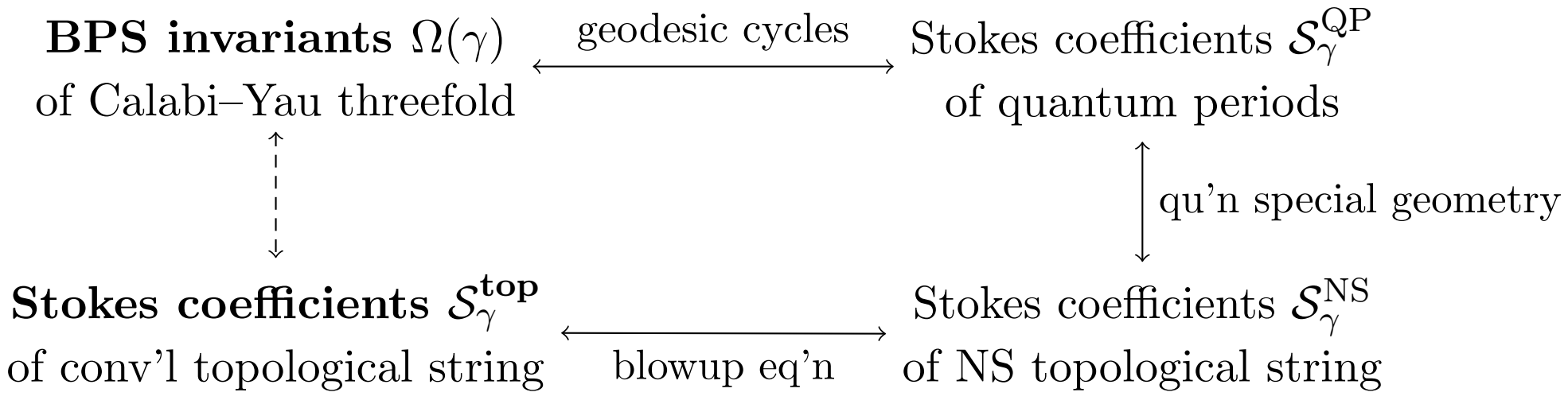

The first argument applies to a special class of topological
string models on local Calabi--Yau threefolds whose mirror
curves are hyperelliptic
\begin{equation}\label{eq:curve}
  \Sigma: \quad y^2 + V(x;u_i) = 0,
\end{equation}
where $u_i$ are complex moduli of the curve. We take the geometric
engineering limit \cite{Katz:1996fh}, so that the model describes a four-
dimensional $\CN=2$ field theory, as in the examples of
Seiberg--Witten theory and $N_f=4$ SQCD that we have mentioned before.
In this case, one can consider the quantisation of the mirror curve,
which is the Schr\"odinger equation of a one-dimensional quantum
mechanical problem\footnote{One can choose one of the moduli and move
  it to the right-hand side of the equation so that it becomes the
  standard form of the Schr\"odinger equation.},
\begin{equation}
  -\hbar^2\pd_x^2\psi(x) + V(x;u_i)\psi(x) = 0.
\end{equation}
The Schr\"odinger equation can be solved using the method of exact WKB
\cite{Voros1983,Delabaere19971:exact,kawai2005algebraic,Delabaere1999,Iwaki:2014ewa}.
In the exact WKB method, the mirror curve $\Sigma$ also plays
an important role and is known as the WKB curve.
For each 1-cycle $\gamma\in H_1(\Sigma,\IZ)$, one can define a quantum
period $\Pi_\gamma(u_i;\hbar)$, which is a power series in $\hbar$.
The coefficients of a quantum period are functions of the moduli
$u_i$,
\begin{equation}
  \Pi_\gamma(u_i;\hbar) = \sum_{n=0}^\infty
  \Pi_\gamma^{(n)}(u_i)\hbar^{2n},
\end{equation}
where the leading coefficient is the classical period
\begin{equation}
  \Pi_\gamma^{(0)}(u_i) = \oint_\gamma y\rd x.
\end{equation}
Furthermore, the coefficients grow factorially, i.e.,
\begin{equation}
  \Pi_\gamma^{(n)}(u_i)\sim (2n)!,
\end{equation}
so that the quantum periods are asymptotic power series, and one can
study their resurgent properties.  It turns out that the Borel
singularities of a quantum period are classical periods associated to
geodesic 1-cycles on the WKB curve \cite{Dillinger:1993}, and the
Stokes automorphism across the Stokes ray passing through the Borel
singularity is famously given by the Delabaere--Dillinger--Pham (DDP) formula
\cite{Dillinger:1993},
\begin{equation}\label{eq:DDP}
  \Pi_\gamma(\hbar) \mapsto \Pi_\gamma + \vev{\gamma,\gamma'}\frac{\hbar}{2\pi}\log\left(1+\sigma_{\gamma'}\re^{2\pi\Pi_{\gamma'}(\hbar)/\hbar}\right)
\end{equation}
% \begin{equation}
%   \mc{V}_\gamma \mapsto \mc{V}_\gamma(1+\sigma_{\gamma'}\mc{V}_{\gamma'})^{\vev{\gamma,\gamma'}}.
% \end{equation}
% where $\mc{V}_\gamma$ is the Voros symbol associed to the quantum
% period
% \begin{equation}
%   \mc{V}_\gamma = \re^{2\pi\Pi_\gamma/\hbar},
% \end{equation}
where $\sigma_{\gamma'} = \pm 1$ depends on the type of the geodesic
1-cycle $\gamma'$.
On the other hand, it was argued in \cite{Klemm:1996bj} that in type
IIA superstring theory compactified on the same Calabi--Yau threefold, the BPS
states are described by non-critical strings wrapping geodesic cycles
on the mirror curve.  It is therefore natural to identify BPS states
in type IIA superstring theory with Borel singularities of quantum
periods.

A complementary field-theoretic explanation was developed by Gaiotto,
Moore, and Neitzke for four-dimensional theories obtained by reducing
the six-dimensional $(2,0)$ theory on a punctured Riemann surface
\cite{Gaiotto:2009hg}. In the rank-two Hitchin-system description,
finite WKB trajectories determine BPS states, while changes of the WKB
triangulation induce Kontsevich--Soibelman transformations of
canonical Darboux coordinates; these transformations are also
of the DDP type.
Furthermore, an isolated saddle trajectory gives the factor associated
with a hypermultiplet of index $\Omega=1$, whereas an annular family
of closed trajectories gives the factor associated with a vector
multiplet of index $\Omega=-2$.  This explains why BPS indices occur
in the jumps of the Darboux coordinates.  Interestingly, in suitable
limits and with appropriate normalization, these Darboux coordinates
are identified with Borel-resummed quantum periods.

Furthermore, on a local Calabi--Yau threefold one can consider the
refined free energy in the NS limit, whose gauge-theory origin is the
quantization proposal of Nekrasov and Shatashvili
\cite{Nekrasov:2009quantization}; see also
\cite{Nekrasov:2002qd,Iqbal:2007ii,Huang:2010kf}. % The quantum
% Seiberg--Witten analysis of \cite{Grassi:2019coc} concerns quantum
% periods, rather than directly establishing the conjecture for the
% self-dual free energy.
The relation between NS free energies and quantum periods, including
their multi-instanton sectors, was studied in \cite{Gu:2022fss}. Using
blowup equations \cite{Nakajima:2005fg,Grassi:2016nnt,Huang:2017mis},
\cite{Gu:2023wum} relates the Stokes constants of NS and unrefined
free energies, up to possible signs, for the charge sectors covered by
that
analysis. % Its scope excludes certain purely electric sectors, so
% it should not be read as an unrestricted equality of all Stokes
% constants.
Together, these results support the proposed BPS interpretation in the
local setting, as illustrated in Fig.~\ref{fig:NS}.
% \end{Historyrevision}

\subsubsection{String moduli space}

From the resurgent properties of the topological string partition
function encoded in \eqref{eq:Znp} one can also work out the Stokes automorphism
of the Zak transform of the partition function, also called the
dual partition function.  For simplicity, let us assume that the
moduli space of the topological string is one-dimensional, with flat
coordinate $t$. The dual partition function is then defined as
\begin{equation}
  \CT(t,\sigma,g_s) = \sum_{n\in\IZ} \re^{2\pi n \sigma/g_s}%
  \CZ(t-\ri g_s n, g_s).
\end{equation}
Across a Stokes ray $d$ that contains multiple Borel singularities
located at $k\CA$ where $\CA_\gamma = c \pd_t \CF_0(t)$, the Stokes
automorphism of the dual partition function reads \cite{Iwaki:2023rst}
\begin{equation}
  \mf{S}_d\CT(t,\sigma,g_s) = \re^{\frac{\CS_\gamma}{2\pi\ri}%
    \Li_2(\re^{2\pi\ri c\sigma/g_s})}
  \CT\left(t+\tfrac{\CS_\gamma}{2\pi}g_s c\log(1-\re^{2\pi
      c\sigma/g_s}),\sigma,g_s\right).
  \label{eq:T-Stokes}
\end{equation}

Two observations are relevant.  First, under the Stokes automorphism, the
flat coordinate $t$ transforms as
\begin{equation}
  t \mapsto t+\tfrac{\CS_\gamma}{2\pi}g_s c\log(1-\re^{2\pi c\sigma/g_s})
\end{equation}
which is similar to the Delabaere--Dillinger--Pham (DDP) formula of
quantum periods \eqref{eq:DDP}, whose Stokes constants are identified
with BPS invariants.

% \begin{GMNrevision}
The logarithmic coordinate transformation has a field-theoretic
counterpart % in the construction of Gaiotto, Moore, and Neitzke
\cite{Gaiotto2008cd,Gaiotto:2009hg}. After compactification on a circle,
the corrected Coulomb-branch metric admits a description in terms of
sectorial Darboux coordinates whose jumps are Kontsevich--Soibelman
symplectomorphisms, with the BPS indices as exponents. Taking logarithms
of these multiplicative jumps gives transformations of the same general
form as the coordinate shift above. % The dilogarithm is the Hamiltonian
% generating an elementary Kontsevich--Soibelman transformation; see
% section~2.2 of \cite{Gaiotto2008cd}. This coordinate-level relationship
% does not by itself identify the topological string dual partition
% function or determine its multiplicative transformation law. The lift
% to the partition function and its interpretation in terms of five-brane
% corrections require the additional input discussed next.
% \end{GMNrevision}

% \begin{Historyrevision}
Secondly, the relation between five-brane instantons and topological
string wave functions was developed in
\cite{Alexandrov:2010fivebrane}. The action of Kontsevich--Soibelman
transformations on the corresponding generalized theta series was
obtained in \cite{Alexandrov:2015theta}.  The comparison with
topological string Stokes automorphisms was made in
\cite{Iwaki:2023rst}. % , and the refined extension was developed in
% \cite{Alexandrov:2023wdj}
In this framework, the dual partition function is related to the
five-brane correction to the hypermultiplet moduli space of type II
string theory, with the appropriate mirror identification. The
correction is described by a section of a line bundle, with local
representatives on overlapping patches. The transformation
\eqref{eq:T-Stokes} matches the corresponding gluing law when the
Stokes constants are identified with BPS invariants.  This
interpretation uses the wave-function and theta-series construction in
addition to the coordinate-level comparison above.
% \end{Historyrevision}

\subsubsection{Wall-crossing}

As we mentioned in section \ref{sc:modularity}, the BPS invariants are
invariant only inside a stability chamber of the moduli space and they
change upon crossing a wall of marginal stability, a phenomenon known
as wall-crossing.  It has been demonstrated explicitly that the Stokes
constants also enjoy the feature of wall-crossing.  For instance, in
the case of topological string on local $\IF_0$ it was shown in 
\cite{Marino:2024yme} that the pattern of Borel singularities and the
associated Stokes constants are different on the two sides of the conifold
point $z = 1/16$.

\begin{figure}
  \centering
  \includegraphics[height=4cm]{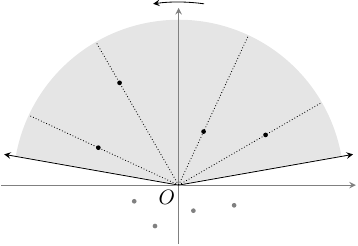}
  \caption{The complex $\IC_Z$ plane.}
  \label{fig:KS}
\end{figure}

The Kontsevich--Soibelman wall-crossing formula controls the
wall-crossing behavior of BPS invariants. We now describe this formula
in more detail.
The Kontsevich--Soibelman (KS) wall-crossing formula is based on the Kontsevich--Soibelman Lie
algebra.
It defines for any charge $\gamma$ in the charge lattice a basis
element $e_\gamma$, and these generators satisfy the Lie algebra
bracket
\begin{equation}%
  [e_\gamma,e_{\gamma'}] =
  (-1)^{\vev{\gamma,\gamma'}}\vev{\gamma,\gamma'}e_{\gamma+\gamma'}.
\end{equation}
Now consider the complex plane $\IC_Z$ of central charges $Z_\gamma$
of BPS states.  In the generic case, two BPS states with coprime
charges $\vev{\gamma,\gamma'} \neq 0$ have different phases in the
central charges.  Then for each \emph{active} ray in the complex plane
$\IC_Z$ that supports a BPS state and its multiples with charges
$n\gamma$ for $n=1,2,3,\ldots$, one can define a Lie group element
\begin{equation}
  K_\gamma = \exp\left(\Omega(\gamma)\sum_{n=1}^\infty\frac{e_{n\gamma}}{n^2}\right).
\end{equation}
For a cone $V$ in the complex plane $\IC_Z$ bounded by two non-active rays
$\rho_1,\rho_2$ as shown in Fig.~\ref{fig:KS}, one can construct a
product of the Lie group elements associated with all the active rays
inside the cone ordered by the phases of these rays,
\begin{equation}
  K_V = \prod_{\gamma:Z_\gamma\in V}^{\rightarrow}K_\gamma,
\end{equation}
Such a global Lie group element can be shown to be a wall-crossing
invariant \cite{Kontsevich:2008fj,Joyce:2008pc} in the sense that when
one moves in the moduli space, even crossing walls of marginal
stability, the global Lie group element is invariant as long as no
central charge of a BPS state enters or exits the cone.

% \begin{GMNrevision}
% Gaiotto, Moore, and Neitzke gave a physical explanation of this
% wall-crossing formula by compactifying a four-dimensional theory on a
% circle \cite{Gaiotto2008cd}. Individual BPS indices can jump at a wall
% of marginal stability, but the corrected hyperk\"ahler metric must remain
% smooth. Their sectorial Darboux coordinates jump by symplectomorphisms
% weighted by BPS indices; consistency when the corresponding rays align
% requires the same ordered product on the two sides of the wall. This
% mechanism is described in section~5.2 of \cite{Gaiotto2008cd}, while
% section~7 formulates the physical argument in terms of Stokes factors
% and isomonodromic deformation. The Hitchin-system construction realizes
% the coordinate jumps through changes of WKB triangulations
% \cite{Gaiotto:2009hg}; see also section~2.3 and equations~(2.10)--(2.11)
% there. These constructions provide a field-theoretic model for the
% sectorial consistency condition in the resurgence discussion below.
% They do not by themselves establish the corresponding identification for
% type II string theory on a general compact Calabi--Yau threefold.
% \end{GMNrevision}

The resurgence picture is analogous.
Recall that the resurgent structure of the topological string partition
function in \eqref{eq:Znp} is described by the following action of alien
derivatives
\begin{equation}
  \dot{\Delta}_\gamma \CZ = \overline{\CS}_\gamma \CD_\gamma \CZ
\end{equation}
where the operator $\CD_\gamma$ is
\begin{equation}
  \CD_\gamma = \frac{1}{2\pi\ri}\left(1+\SD_\gamma\right)\re^{-\SD_\gamma}
\end{equation}
with $\SD_\gamma$ given by \eqref{eq:SD}. The Stokes constants
are defined in terms of the primitive Stokes constants by \eqref{eq:multcover}.
For a Stokes ray that supports the Borel singularities with charge
$\gamma$ and its multiples, the Stokes automorphism is then
\begin{equation}
  \mr{S}_\gamma = \exp\left(\CS_\gamma \sum_{n=1}^\infty\frac{\mc{D}_{n\gamma}}{n^2}\right).
\end{equation}
Furthermore, for a cone $V$ in the Borel plane bounded by two rays
$\rho_1,\rho_2$ which are not Stokes rays, one can define the
sectorial Stokes automorphism
\begin{equation}
  \mr{S}_V = \prod_{\gamma: \CA_\gamma\in V}^{\rightarrow} \mr{S}_\gamma,
\end{equation}
% \begin{GMNrevision}
% The analogous sectorial consistency condition is that
% $\mr{S}_V$ remain invariant under deformations of the moduli for
% which no Borel singularity enters or exits the cone. Establishing this
% invariance requires compatible sectorial analytic continuation, or an
% appropriate isomonodromic description; continuity of the resummed
% partition functions alone is not sufficient. The construction of
% \cite{Gaiotto2008cd,Gaiotto:2009hg} explains this mechanism for the
% field-theoretic Darboux coordinates.
% \end{GMNrevision}
Moreover,
one can show that the alien derivatives satisfy the same Lie bracket
relations as the KS algebra \cite{Douaud:2026qfo,Guo:2026xpg}
\begin{equation}%
  [\CD_\gamma,\CD_{\gamma'}] =
  (-1)^{\vev{\gamma,\gamma'}}\vev{\gamma,\gamma'}\CD_{\gamma+\gamma'}.
\end{equation}
% \begin{GMNrevision}
% The realization by the operators $\CD_\gamma$ is additional
% topological string input, rather than a result of
% \cite{Gaiotto2008cd,Gaiotto:2009hg}; it requires its own derivation and
% compatible normalization conventions. Sharing the Lie algebra alone
% also does not fix the spectrum of Stokes constants: the comparison
% requires the chamber and the charge-dependent data.
% \end{GMNrevision}

It is therefore natural to identify the sectorial Stokes automorphism with
the KS wall-crossing invariant, the alien derivatives with the KS Lie
algebra generators, and the primitive Stokes constants
$\CS_\gamma$ with the BPS invariants $\Omega(\gamma)$.

% \subsection{Relation with TS/ST correspondence}

% \jjj{Move to evidences?}

% \subsection{Generalization: Defects and refinements}

% The conjectures that relate the resurgent structure of topological
% string free energy and BPS states of type II superstring have been
% generalized in various directions.

\section{Prospects}

In this paper, we have briefly reviewed the conjectural connection between
the resurgent structure of topological string free energy and BPS
states of type II superstring theory, which are stable D-brane bound states.
The Borel singularities of the topological string free energy
correspond to central charges of the BPS states, while the Stokes
constants are identified with the BPS invariants.  Both the Stokes
constants and the BPS invariants have wall-crossing behavior and they
seem to be controlled by the same Kontsevich--Soibelman wall-crossing
formalism.

These conjectures have already been generalized in various directions.
For instance, studies have addressed the resurgence of free energies
in real topological string theory \cite{Marino:2023nem}, fully refined
topological string theory \cite{Alexandrov:2023wdj}, and backgrounds
with Wilson-loop insertions \cite{Gu:2024wag}.

Many open questions remain.  The most obvious open question is
whether the supporting arguments discussed in section
\ref{sc:arguments} can be developed into a rigorous proof, or whether new
ideas are required.

Furthermore, given the conjectural roles of Borel singularities and
Stokes constants in the correspondence with BPS states in type II
superstring, an important missing element is the role of the
non-perturbative series. Is it related to the partition function with
a D-brane insertion?

Moreover, resurgence theory need not capture every non-perturbative
contribution.  For instance, in the double-well or periodic cosine
model of quantum mechanics, resurgence theory applied to the
perturbative energy series misses the 1-instanton contribution or the
Bloch-angle dependence, respectively.  In the case of topological
string, does resurgence capture all possible non-perturbative
contributions?  What about contributions of the order
$\mc{O}(\re^{-1/g_s^2})$ associated to Neveu--Schwarz five-branes
(NS5-branes)? Indications that this type of non-perturbative
contribution exists in topological string theory have been found
\cite{Couso-Santamaria2017a}.

This is also related to the convergence problem of instanton sums.
Typically, along a Stokes ray of the topological string free energy,
there are multiple Borel singularities corresponding to
multi-instanton contributions.  To formulate the Stokes
automorphism across the Stokes ray, one has to sum over all the
multi-instanton contributions.  Is the sum convergent?  There are
indications that the multi-instanton series is asymptotic, as the BPS
invariants or the Stokes constants grow as $\mc{O}(\re^{n^2})$, and
although there are methods to resum all the multi-instanton
contributions, ambiguities of the order $\mc{O}(\re^{-1/g_s^2})$ arise
\cite{Marino:Strings2026}.

One possible way to explore these two related problems is to formulate
and solve the Riemann--Hilbert problem associated to the Stokes
automorphisms.  This is the approach that has been taken to study the
BPS invariants of 4d $\mc{N}=2$ field theories \cite{Gaiotto2008cd},
but extending the problem to five-dimensional (5d) $\mc{N}=1$ field
theories or to the topological string setting is significantly more
difficult.  See also the related work in
\cite{Bridgeland:2016RH,Bridgeland:2017conifold}.

Finally, it would be very interesting to study the resurgent
properties of open-string partition functions in topological string.
One can consider the NS partition function or the conventional
or self-dual partition function, which are related by the blowup
equations \cite{Kim:2025qaf}, or the fully refined partition
function, in the spirit of \cite{Alexandrov:2023wdj}.  % \begin{Historyrevision}
For open strings, the difference-equation and exponential-network
analyses of \cite{Grassi:2022zuk,Alim:2022curves} treat quantum mirror
curves and brane amplitudes in simple geometries. These results should
be distinguished from a general analysis of the self-dual open-string
partition function. For the NS open-string sector, exact WKB studies
of quantum-curve wave functions provide a natural starting point
\cite{Grassi:2021wpw,Nikolaev:2024Geometry}.
% \end{Historyrevision} 

%%%%%%%%%%%%%%%%%%%%%%%%%%%%%%%%%%%%%%%%%%%%%%%%%%%%%%%%%%%%%%%%%%%%%%%%%%%%%%%%%% 
\printindex

% Bibliography	
% \bibliographystyle{\bibtexstylefile}
\bibliographystyle{amsmod}
% Version 6 uses the self-contained ATMP-report-v6.bib database.
%\input{ATMP-report-v4.bbl}
\bibliography{ATMP-report}

\end{document}